**High-pressure elastic properties of $GeO_2$ polymorphs up to 120 GPa**

Gulshan Kumar[1,*], Sumit Ghosh[1], Sharad Babu Pillai[1], and Rajkrishna Dutta[2]
[1)] *Department of Earth Sciences, IIT Gandhinagar, Gujarat 382355, India*
[2)] *Department of Geosciences, Princeton University, Princeton, New Jersey 08544, USA*

**Abstract**

We systematically investigated the phase stability, pressure dependence of the elastic properties, and elastic anisotropy of four $GeO_2$ polymorphs: rutile-, $CaCl_2$-, $\alpha$-$PbO_2$-, and pyrite-type phases using theoretical calculations based on density functional theory. The elastic constants were calculated at regular pressure intervals within the stability fields of each phase, as determined by static enthalpy calculations. We further employed a strain-coupled Landau free-energy expansion to describe the pressure evolution of the elastic response associated with the rutile- to $CaCl_2$-type transition and elucidate the origin of the elastic softening near the transition. The rutile- to $CaCl_2$-type phase transition is consistent with a Landau-type second-order transition. As the transition is approached, elastic softening develops in the rutile-type phase, resulting in anomalous pressure dependence of the bulk and shear moduli. The calculated elastic-wave anisotropy increases markedly near the transition, primarily due to the rapid reduction in shear-wave velocity, reaching a maximum of approximately 190% at 20 GPa. Following the transition, the anisotropy decreases sharply in the $CaCl_2$-type phase and exhibits discontinuity at the $CaCl_2$-type/$\alpha$-$PbO_2$-type phase boundary. The higher-pressure $\alpha$-$PbO_2$- and pyrite-type phases exhibit comparatively weak pressure dependence of anisotropy, with small discontinuities at their respective phase boundaries. The pyrite-type phase has the lowest anisotropy with $A_P \approx 4\%$ and $A_S \approx 12\%$ at 120 GPa, consistent with its high-symmetry cubic structure and nearly isotropic elastic-wave propagation.

*Corresponding author: Gulshan Kumar (kumargulshan@iitgn.ac.in)

## Introduction:

Germanium dioxide ($GeO_2$) has been widely studied because it is a structural and chemical analog of silicon dioxide ($SiO_2$) and displays several pressure-induced phase transitions [1]. It is also of significant interest because of its important optical, electronic, and piezoelectric applications [2–12]. Therefore, a clear understanding of its mechanical stability and high-pressure behavior is of considerable importance in both geosciences and condensed matter physics. First-principles calculations based on density functional theory (DFT) provide a powerful approach for understanding the structural, electronic, elastic, and optical properties of materials [13–17].

At ambient conditions, crystalline $GeO_2$ crystallizes in a rutile-like tetragonal phase with space group $P4_2/mnm$ [18]. With increasing pressure, the rutile-type phase first transforms into the $CaCl_2$-type phase (26 GPa, *Pnnm*), followed by transitions to the α-$PbO_2$-type (36 GPa, *Pbcn*) phase and finally to a pyrite-type phase (65 GPa, $Pa\bar{3}$) [19–21]. This transition sequence is consistent with the phase transitions in other group-14 dioxides, such as $SiO_2$ [22], $SnO_2$ [23,24], and $PbO_2$ [25,26].

Raman spectroscopic studies, supported by both experimental observations [27] and theoretical calculations [28], suggest that the rutile- to $CaCl_2$-type phase transition in $GeO_2$ is a second-order ferroelastic transition. This transition is associated with the softening of the Raman-active $B_{1g}$ mode and its coupling with acoustic modes, resulting in a strong nonlinear response of the elastic constants [29–31]. Ferroelastic phase transitions of this type are commonly interpreted within the framework of Landau theory, in which the transition is described by a symmetry-breaking order parameter that couples to spontaneous strain [32]. The other two transitions differ from the rutile- to $CaCl_2$-type transformation because they are reconstructive and first-order in character [20]. The $CaCl_2$-type → α-$PbO_2$-type → pyrite-type transitions involve a rearrangement of the $GeO_6$ octahedral framework, which is associated with a finite volume change [20,21,33,34].

To understand the mechanism of these pressure-induced phase transitions, it is therefore important to study the elastic properties of these phases as pressure increases. Experimental [35] and theoretical [20,36] investigations of the elastic constants have primarily focused on the low-pressure rutile- and $CaCl_2$-type phases. However, studies of the elastic properties of $GeO_2$ at high pressures (>30 GPa) remain limited.

In this study, we establish the pressure evolution of the full elastic tensor of rutile-, $CaCl_2$-, α-$PbO_2$-, and pyrite-type $GeO_2$ from 0 to 120 GPa, quantitatively analyze the coupling between the order parameter and strain within a Landau formalism to explain elastic softening across the rutile-to-$CaCl_2$ transition, and characterize changes in elastic-wave anisotropy across all three phase boundaries.

**Computational method:**

All computations were performed using plane-wave density functional theory (DFT) [37,38], as implemented in the Quantum ESPRESSO package [39,40]. The local density approximation (LDA) was used for the exchange-correlation functional. Electron-ion interactions were treated using Vanderbilt's ultrasoft pseudopotentials [41]. A kinetic energy cutoff of 110 Ry was used for the basis set in all phases. The reciprocal space was sampled using a Monkhorst–Pack 11 × 11 × 11 k-point grid [42]. Both lattice parameters and atomic positions were optimized at each pressure step using the Broyden-Fletcher-Goldfarb-Shanno (BFGS) algorithm [43]. Geometry optimizations were considered complete when the atomic forces were less than $2\times10^{-4}$ Ry/Bohr. Density functional perturbation theory (DFPT) was used to calculate the frequencies of the $B_{1g}$ (soft) mode and $A_g$ (hard) mode for the rutile-type and $CaCl_2$-type phases at the gamma ($\Gamma$) point [q = (0,0,0)].

To calculate the elastic constants ($C_{ij}$) and elastic moduli, the equilibrium structures at each pressure were subjected to a series of small deformations. The maximum absolute value of the Lagrangian strain set to $3\times10^{-2}$, producing 19 distorted structures using the deformation types implemented in the ElaStic code [44]. Internal degrees of freedom were optimized, and the total energy of all the deformed structures was obtained using Quantum ESPRESSO [39,40]. To determine the second derivative of the energy with respect to Lagrangian strain at equilibrium, the calculated energy–strain data were fitted using a polynomial function. The coefficient of the quadratic term in the best-fitting polynomial can be expressed as a linear combination of the second-order elastic constants. This results in a set of linear equations for each deformed structure, which are subsequently solved using a least-squares fitting method to determine the second-order elastic constants ($C_{ij}$). A pressure correction was applied to the calculated elastic constants to account for the finite external pressure P, under which the calculations were performed [45].

## Results and Discussion:

a. Structural stability and phase transitions:

The structural properties of the rutile-, $CaCl_2$-, $\alpha$-$PbO_2$-, and pyrite-type $GeO_2$ polymorphs (Figure 1) were investigated over the pressure range of 0-120 GPa at regular pressure intervals of 5 GPa (10 GPa above 100 GPa), with additional calculations at 28 and 32 GPa to better resolve the narrow stability field of the $CaCl_2$-type phase. The optimized lattice parameters and unit-cell volumes are summarized in Table I and compared with existing theoretical and experimental values. The rutile-type ($P4_2/mnm$) structure of $GeO_2$ consists of edge-sharing $GeO_6$ octahedra arranged in straight chains parallel to the tetragonal $c$-axis, with adjacent chains linked by corner-sharing oxygen atoms in the $ab$-plane [46]. In the $CaCl_2$-type ($Pnnm$) phase, the $GeO_6$ octahedra retain the same connectivity as in the rutile structure but differ by an orthorhombic distortion arising from tilting and rotation of the octahedral chains along the $c$-axis [46,47]. In contrast to the straight octahedral chains of the rutile and $CaCl_2$-type structures, the $\alpha$-$PbO_2$-type phase features zigzag chains of edge-sharing octahedra, together with a significant displacement of the Ge cations and a loss of the original chain alignment along the $c$-axis [34,47]. The pyrite-type ($Pa\bar{3}$) phase represents the densest $GeO_2$ polymorph. In this structure, $GeO_6$ octahedra form a three-dimensional framework connected exclusively through corner sharing, with no edge sharing between adjacent octahedra [47]. Figure 2 shows the enthalpy ($H$) difference of each structure with respect to the rutile-type phase as a function of pressure at 0 K. The phase transition pressures, $P_T$ were determined by identifying the intersections of the enthalpy–pressure curves of the competing phases and interpolating between the calculated points. At each pressure, the phase with the lowest enthalpy among the polymorphs is thermodynamically stable at 0 K.

Our computations show that the tetragonal rutile- to orthorhombic $CaCl_2$-type phase transition occurs at ~24 GPa. Although this value is higher than the transition pressure predicted by DFT calculations (~19 GPa) [20], it is in excellent agreement with existing experimental observations (25–28 GPa) [27,46]. It is also in good agreement with recent theoretical studies (22.6–25.2 GPa) [28]. The $CaCl_2$-type phase of $GeO_2$ is expected to transform into the orthorhombic $\alpha$-$PbO_2$-type structure at ~33 GPa. This value is lower than that reported in laser-heated diamond anvil cell (LHDAC) experiments [34], where the $\alpha$-$PbO_2$-type phase was identified as the stable post-$CaCl_2$ polymorph above 44 GPa. Coexistence of $CaCl_2$- and $\alpha$-$PbO_2$-type $GeO_2$ was reported to persist up to 60 GPa. Our predicted transition pressure is in good agreement with existing first-principles calculations, which reported the $CaCl_2$- → $\alpha$-$PbO_2$-type transition over a pressure range of 30.5–36 GPa [20,21].

The $\alpha$-$PbO_2$-type phase is predicted to transform into the pyrite-type structure at 63 GPa. This transition pressure is lower than that reported in the LHDAC experiments [48], where the pyrite-type phase was observed above 90 GPa. Although our calculated transition pressure is significantly lower than the experimentally determined value, it is in good agreement with previous theoretical predictions, which reported transition pressures in the range of 59–69.5 GPa [20,21]. The differences in the reported transition pressures may arise from variations in experimental conditions, including the pressure-transmitting medium, pressure calibrant, degree of hydrostaticity, thermal gradients within the DAC [49], as well as computational choices, such as the exchange-correlation functional and pseudopotentials. Our 0 K calculations use the LDA exchange–correlation functional, which tends to overbind and favor denser phases. Thermal and zero-point contributions are not included in our calculations. Moreover, the experiments observed the $\alpha$-$PbO_2$-type phase at ~60 GPa and the pyrite-type phase at ~100 GPa, with no measurements between these pressures; thus, a lower experimental transition pressure cannot be ruled out. The present work provides a complete and internally consistent dataset of the structural and elastic properties of all four $GeO_2$ polymorphs obtained within a single computational framework. Figure 3 shows the volume per formula unit as a function of pressure. Volume change across the rutile-to $CaCl_2$-type phase boundary is negligible, consistent with a second-order phase transition. The phase transition is characterized by the splitting of the equivalent $a$-axis of the tetragonal rutile-type structure into inequivalent $a$- and $b$-axes in the $CaCl_2$-type phase, reflecting an orthorhombic distortion [28]. The $CaCl_2$- → $\alpha$-$PbO_2$- → pyrite-type phase transitions are characterized by significant atomic rearrangements and finite volume discontinuities (1.17% and 4.70%, respectively), consistent with their first-order nature.

a. Elastic constants:

Elastic constants derived from DFT energy–strain calculations directly capture the microscopic elastic response and allow the analysis to be extended to the higher-pressure polymorphs. We have therefore calculated the pressure-dependent single-crystal elastic constants for all four phases within their stability range. Rutile-type $GeO_2$ has tetragonal symmetry and is characterized by six independent elastic constants ($C_{11}$, $C_{12}$, $C_{13}$, $C_{33}$, $C_{44}$, and $C_{66}$). The $CaCl_2$- and $\alpha$-$PbO_2$-type phases exhibit orthorhombic symmetry and are described by nine independent elastic constants ($C_{11}$, $C_{22}$, $C_{33}$, $C_{12}$, $C_{13}$, $C_{23}$, $C_{44}$, $C_{55}$, and $C_{66}$). The pyrite-type phase crystallizes in a cubic structure and is characterized by three independent elastic constants ($C_{11}$, $C_{12}$, and $C_{44}$). The pressure dependence of the elastic constants is shown in Figure 4.

In the rutile-type phase, $C_{33}$ is the largest elastic constant and increases strongly with pressure, indicating increasing stiffness along the $c$-axis. $C_{11}$ also increases with pressure but remains lower than $C_{33}$, consistent with high-pressure X-ray diffraction results showing lower compressibility along the $c$-axis than along the $a$-axis [28,46]. The relatively lower values of $C_{44}$ indicate comparatively weaker resistance to shear deformation, whereas the smaller $C_{12}$ and $C_{13}$ values reflect weaker coupling between normal strains along different crystallographic directions. While $C_{44}$ and $C_{13}$ show only weak pressure dependence, $C_{12}$ and $C_{66}$ increase more significantly with pressure. The increase in $C_{66}$ suggests enhanced resistance to shear deformation within the basal plane as the rutile- to $CaCl_2$-type transition is approached. Table II compares the elastic constants obtained in the present study with available experimental and theoretical results at selected pressures. For rutile-type $GeO_2$ at ambient pressure, our calculated values are in good agreement with experimental measurements [35] and previous theoretical results [20].

In the orthorhombic $CaCl_2$-type phase, $C_{33}$ remains the largest elastic constant, indicating high stiffness along the *c*-axis. $C_{11}$ increases rapidly with pressure and becomes significantly larger than $C_{22}$, highlighting the strong elastic anisotropy of the orthorhombic structure. The splitting of $C_{11}$, $C_{22}$, and $C_{33}$ reflects the reduction in symmetry from tetragonal rutile to orthorhombic $CaCl_2$-type $GeO_2$. In contrast, $C_{12}$ and $C_{13}$ decrease slightly with pressure, indicating a weakening of the coupling between the corresponding normal strains. The shear elastic constants $C_{44}$, $C_{55}$, and $C_{66}$ show relatively weak pressure dependence over the investigated pressure range.

In the $\alpha$-$PbO_2$-type phase, all the elastic constants increase nearly linearly with pressure between 35 and 60 GPa, indicating progressive mechanical stiffening under pressure. The principal normal elastic constants follow the order $C_{33}>C_{22}>C_{11}$, indicating that the structure is stiffest along the *c*-axis, followed by the *b*- and *a*-axes. Compared with the $CaCl_2$-type phase, the pressure dependence is more systematic. The increases in $C_{12}$, $C_{13}$, and $C_{23}$ indicate stronger coupling between the corresponding normal strain directions, while $C_{44}$, $C_{55}$, and $C_{66}$ remain lower than the normal elastic constants, reflecting the comparatively lower resistance to the corresponding shear deformations. For the $CaCl_2$- and $\alpha$-$PbO_2$-type phases, our elastic constants at 30 and 40 GPa are broadly consistent with the previous first-principles calculations [36].

The pyrite-type phase shows the highest elastic constants among the four $GeO_2$ polymorphs. $C_{11}$ is exceptionally large and increases from ~917 GPa at 65 GPa to ~1206 GPa at 120 GPa, indicating a very high longitudinal stiffness and strong resistance to deformation along the corresponding crystallographic direction. $C_{12}$ and $C_{44}$ also increase steadily with pressure but remain substantially smaller than $C_{11}$. This difference indicates weaker coupling between the corresponding normal strains and a lower resistance to shear deformation relative to the longitudinal response represented by $C_{11}$. Apart from DFT calculations of elastic constants at specific pressures [36], pressure-dependent single-crystal elastic constants of the $\alpha$-$PbO_2$- and pyrite-type phases of $GeO_2$ have not been reported, and no experimental data are available to the best of our knowledge. The present results therefore provide a useful reference dataset for understanding the elastic properties and pressure evolution of these high-pressure $GeO_2$ polymorphs.

The mechanical stability of the $GeO_2$ polymorphs was assessed using the appropriate pressure-dependent Born stability criteria at each pressure [50]. In the rutile-type phase, $C_{11}$ initially increases with pressure, but its rate of increase slows down upon further compression, whereas $C_{12}$ increases rapidly with pressure. Consequently, the tetragonal stability condition $(C_{11}-C_{12}-2P)>0$ [50] is violated above 20 GPa, indicating a shear instability associated with the rutile- to $CaCl_2$-type transition. This is consistent with the theoretical study [20], which reported the instability at ~19 GPa. For the $CaCl_2$-type phase, most elastic constants, including $C_{11}$, $C_{22}$, $C_{33}$, $C_{23}$, $C_{44}$, $C_{55}$, and $C_{66}$, increase with pressure, although $C_{44}$ is only weakly pressure sensitive. In contrast, $C_{12}$ and $C_{13}$ decrease with compression, indicating weakening of selected interaxial strain couplings. Nevertheless, the calculated constants satisfy the orthorhombic Born stability criteria throughout the $CaCl_2$ stability field, confirming its mechanical stability [50]. The $\alpha$-$PbO_2$-type phase also satisfies the orthorhombic stability criteria up to its transition pressure of 63 GPa. For the cubic pyrite-type phase, the conditions $(C_{11}-C_{12}-2P)>0$; $C_{44}-P>0$; and $(C_{11}+2C_{12}+P)>0$ remain fulfilled over the studied pressure range, indicating that this dense high-pressure phase is mechanically stable under compression [50].

Figure 5 shows the pressure dependence of the Hill average bulk modulus (*K*) and shear modulus (*G*), together with Reuss and Voigt moduli for the rutile-, $CaCl_2$-, $\alpha$-$PbO_2$- and pyrite-type phases of $GeO_2$. Across the rutile- → $CaCl_2$-type transition, the Hill-average bulk modulus decreases by ~9%, whereas it increases by ~10% and ~5.5% across the $CaCl_2$- → $\alpha$-$PbO_2$-type and $\alpha$-$PbO_2$ → pyrite-type transitions, respectively. On the other hand, the Hill-average shear modulus increases by ~51% at the rutile- → $CaCl_2$-type transition, decreases by ~9% at the $CaCl_2$- → $\alpha$-$PbO_2$-type transition and increases again by ~22% upon transformation to the pyrite-type phase. In all four phases, *K* is significantly larger than *G*, indicating that $GeO_2$ resists volume compression more strongly than shear deformation. In the rutile phase, *G* remains nearly pressure independent and shows a slight decrease close to the transition pressure, indicating shear softening as the structure approaches the rutile → $CaCl_2$ transformation. In the other three high-pressure phases, both the bulk and shear moduli increase approximately linearly with pressure, indicating progressive elastic stiffening upon compression.

b. Landau model:

The rutile- to $CaCl_2$-type transition in $GeO_2$ is a second-order ferroelastic transition characterized by the absence of volume discontinuity and a continuous evolution of the structural order parameter [27,28]. The transition is driven by the softening of the symmetry-breaking elastic constant combination ($C_{11}$-$C_{12}$) as the transition pressure is approached [51,52]. While DFT calculations quantitatively capture the phase transition and the pressure dependence of the elastic constants, they do not explicitly describe the mechanism of elastic softening in terms of the coupling between the symmetry-breaking order parameter and strain. In contrast, Landau free-energy expansion theory (see supplementary material for details) provides a classical framework for describing the coupling between the order parameter (Q) and spontaneous strain [32,51]. The Landau framework establishes a connection between microscopic lattice instability and macroscopic elastic anomaly. In the Landau formulation, the order parameter, Q is linearly coupled to the symmetry-breaking strain $(e_1 - e_2)$. Q remains zero in the high-symmetry structure (rutile-type) and attains a finite positive value in the low-symmetry phase ($CaCl_2$-type) above a strain-derived critical pressure, $P_S$ (Figure S1). $P_S$ is the pressure at which, within the Landau model, the rutile-type structure becomes unstable against the symmetry-breaking strain. $P_S$ =18.3 GPa is lower than the enthalpy-driven phase-transition pressure, i.e., $P_T$ = 24 GPa, because $P_S$ represents the bare critical pressure associated with spontaneous strain. Coupling between strain and the order parameter renormalizes the Landau coefficients, including the effective fourth-order coefficient of the order parameter [32,53]. The enthalpy transition pressure corresponds to the thermodynamic equilibrium phase boundary, where the $CaCl_2$-type phase becomes energetically favored over the rutile-type phase, as shown in Figure 2. The distinction between $P_S$ and the enthalpy crossover pressure, therefore, reflects the difference between the mechanical instability of the rutile-type structure and the thermodynamic stability of the two competing phases.

To quantify the energetic contribution associated with the Landau transition, we evaluated the excess Gibbs free energy, $\Delta G_{excess}$ using a strain-coupled Landau free energy expansion (Eq. S1 of the supplementary material). The total Gibbs free energy can be expressed as:

$$G_{total} = G_{bg} + \Delta G_{excess}$$

where $G_{bg}$ represents the smooth background contribution of the rutile-type phase [32]. $\Delta G_{excess}$ represents the excess Gibbs free energy contribution associated with the symmetry-breaking order parameter and its coupling to spontaneous strain. $\Delta G_{excess}$ is zero in the rutile-type structure because both the order parameter $Q$ and the spontaneous strains $(e_1 - e_2)$ is zero. Once the symmetry-breaking distortion develops above the critical pressure, $Q$ becomes non-zero and $\Delta G_{excess}$ turns negative, indicating that the symmetry-broken state is energetically favored relative to the smooth rutile-type background. Its magnitude increases continuously with pressure, reflecting the progressive stabilization of the $CaCl_2$-type phase and the increasing energetic contribution of the order-parameter and strain coupling to the phase transition, as shown in Figure 6.

The bare elastic constants used in the Landau formalism represent the elastic stiffness of the high-symmetry tetragonal parent phase in the absence of any symmetry-breaking distortion. These quantities are pressure-dependent because compression affects the intrinsic modulus of the parent structure [32,51–59]. The symmetry-breaking combination ($C_{11} - C_{12}$) softens as the transition pressure is reached, while its bare counterpart ($C^o_{11} - C^o_{12}$) is treated as pressure-independent within the Landau formulation (see supplementary material) and represents the intrinsic elastic response of the high-symmetry phase in the absence of $Q$-strain coupling.

Using the Landau coefficients listed in Table III and the corresponding derivatives of the free energy given in Equation S11 of the Supplementary Material, we calculated the pressure dependence of the elastic constants. The pressure dependence of the elastic constants predicted by the Landau model is shown in Figure 7a. $C_{66}$ increases linearly with pressure in both the rutile- and $CaCl_2$-type phases, whereas $C_{33}$ and $C_{12}$ show a slight decrease near the transition before increasing again with further compression. The tetragonal-to-orthorhombic symmetry lowering lifts the equivalence of several elastic constants, resulting in the splitting of the corresponding elastic constants as shown in Figure 7a. The tetragonal $C_{11}$ splits into $C_{11}$ and $C_{22}$, both of which increase with pressure. Similarly, the tetragonal $C_{13}$ splits into $C_{13}$ and $C_{23}$, where $C_{13}$ increases while $C_{23}$ decreases at the critical pressure. Following the transition, both elastic constants remain nearly pressure independent. The tetragonal $C_{44}$ splits into the orthorhombic $C_{44}$ and $C_{55}$, both of which increase linearly with pressure. The most significant feature is the pronounced softening of the symmetry-breaking elastic combination $C_{11}$-$C_{12}$ in the tetragonal rutile phase, which decreases with pressure and approaches zero at the critical pressure ($P_S$), indicating the loss of stability against the symmetry-breaking distortion. In contrast, the corresponding orthorhombic combination $0.5(C_{11} + C_{22} - 2C_{12})$ increases beyond the transition pressure, indicating the progressive recovery of stability against the symmetry-breaking distortion. This behavior represents the mechanical signature of the ferroelastic instability and reflects coupling between the order parameter and the spontaneous strain. Within the Landau formalism, $C_{12} = C^0_{12} + \lambda_2^2\chi$, $C_{11} = C^0_{11} - \lambda_2^2\chi$ where $\lambda_2$ is the strain-order parameter coupling coefficient and $\chi$ is the susceptibility of the order parameter (see supplementary material for more details). As the transition is approached, increasing susceptibility enhances the strain-order parameter coupling, progressively reducing the restoring force against the tetragonal shear deformation. The continuous softening of $(C_{11} - C_{12})$ towards the critical pressure is consistent with a second-order phase transition..

To understand the macroscopic elastic response, the aggregate bulk modulus ($K$) and shear modulus ($G$) were calculated from the full elastic tensor using the Voigt-Reuss-Hill averaging methods [60]. Figure 7b shows the change in the elastic moduli with pressure. The bulk modulus increases with pressure in both the rutile- and $CaCl_2$-type phases, reflecting the progressive stiffening of the lattice under compression. Near the transition pressure, the Reuss and Hill bulk moduli exhibit a pronounced discontinuity, while the Voigt bulk modulus shows a relatively weak decrease. In the rutile-type phase, the Reuss and Hill shear moduli decrease with pressure. The Reuss shear modulus exhibits the most pronounced softening, approaching zero at the critical pressure, while the Hill average also decreases significantly but remains finite. In contrast, the Voigt shear modulus shows relatively weak pressure dependence and does not exhibit a sharp decrease at the transition pressure. Following the transition to the $CaCl_2$-type phase, all three shear moduli increase with pressure, reflecting the progressive increase in resistance to shear deformation under compression.

c. Elastic anisotropy:

The elastic constants and elastic softening associated with the rutile-type to $CaCl_2$-type ferroelastic transition are expected to strongly affect the directional elastic wave propagation. To understand this velocity anisotropy at high pressure and quantify its evolution across the transition, directional compressional-wave (P-wave) and shear-wave (S-wave) velocities were calculated using the Christoffel matrix formulation [61]: $\Gamma_{ik} = C_{ijkl} n_k n_l$ , where $\Gamma_{ik}$ is the Christoffel matrix, $\hat{n} = (n_1, n_2, n_3)$ is the unit vector describing propagation direction, and $C_{ijkl}$ are the elastic constants. The wave velocities are obtained by solving the Christoffel eigenvalue equation, given by: $\Gamma_{ik} u_k = \rho V^2 u_k$, where $u_k$ is the polarisation wave vector, $V$ is the wave velocity, and ρ is the density at the given pressure. The three eigenvalues yield one longitudinal ($V_P$) and two transverse modes ($V_{S1}$ and $V_{S2}$) for a given propagation direction. After obtaining the full set of directional velocities, we identify the maximum and minimum values ($V_{max}$ and $V_{min}$) for each wave mode and use them to evaluate directional variations using the following relation [62,63]:

$$A_P = \frac{V_{P,max} - V_{P,min}}{V_{P,agg}} \times 100 \, , A_S = \frac{V_{S,max} - V_{S,min}}{V_{S,agg}} \times 100. \quad (1)$$

where $V_{P,agg}$ and $V_{S,agg}$ are the aggregate P-wave and S-wave velocities, respectively; AP and AS are compressional and shear wave anisotropy, respectively.

Figure 8 shows the compressional and shear wave anisotropy for all four stable $GeO_2$ phases. Among these phases, the rutile-type structure exhibits the strongest elastic anisotropy. In this phase, $A_S$ rises sharply from ~71% at ambient pressure to ~190% at 20 GPa near the rutile- to $CaCl_2$-type transition pressure, while $A_P$ increases from ~28% to ~33% with pressure during the same stable pressure range. The sharp increase near the phase boundary is associated with the softening of ($C_{11} - C_{12}$) as the transition is approached. This softening reduces the elastic restoring forces in specific deformation modes, thereby enhancing the directional variation of the P- and S-wave velocities [31]. Elastic-wave anisotropy is governed by the full elastic stiffness tensor through the Christoffel matrix, $C_{44}$, $C_{66}$, and other elastic constants and their coupling terms, which contribute differently depending on the crystallographic propagation directions. In the rutile-type phase, the fastest and slowest P-wave directions are [001] and [100], and the fastest and slowest S-wave directions are [100] and [110]. In the $CaCl_2$-type phase, the fastest P-wave direction is [110], and the slowest is [010], while the S-wave directions are unchanged: fastest along [100] and slowest close to [110]. Following transformation to the $CaCl_2$-type phase, both anisotropies decrease with pressure, with $A_S$ dropping from ~89% at ~25 GPa to ~63% at 32 GPa close to the second phase transition boundary (~33 GPa), while $A_P$ decreases from ~32% to ~28% in the same range.

The α-$PbO_2$-type phase shows a significant drop in elastic wave anisotropy and remains nearly pressure-independent, with $A_S \approx 28\%$ and $A_P \approx 11\%$ across its stable range (33–63 GPa). Finally, the pyrite-type phase, stable above ~63 GPa, shows the lowest anisotropy among all phases, with $A_S$ slowly increasing only from ~8% to ~12% and $A_P$ from ~3% to ~4% up to 120 GPa. The relatively weak directional dependence is consistent with its high-symmetry cubic structure. Although the pyrite-type phase has cubic symmetry, its elastic-wave anisotropy is not necessarily zero. Complete elastic isotropy is achieved only when the Zener anisotropy ratio, $A_Z = 2C_{44}/(C_{11}-C_{12})$, equals unity. Thus, deviations of $A_Z$ from 1 result in finite directional variations of the P- and S-wave velocities, consistent with the small but nonzero values of $A_P$ and $A_S$ calculated for the pyrite-type phase [64].

Understanding the high-pressure behavior of $GeO_2$ provides insights into pressure-induced phase transitions, elastic behavior, and elastic anisotropy in $AO_2$ (A = Si, Ge, Sn, Ti, etc.) compounds. $GeO_2$ shows a similar sequence of phase transitions as $SiO_2$, but at significantly lower pressures [33]. Therefore, $GeO_2$ serves as a low-pressure analog of $SiO_2$, allowing the investigation of pressure-induced phase transitions and mechanical properties under compression under controlled, experimentally accessible conditions. Silica is the most abundant component of the mantle and the crust in the Earth. These insights can therefore help improve our understanding of the elastic behavior of $SiO_2$ and related oxides at conditions relevant to the deep Earth or exoplanetary interiors. Our high-pressure data on the equation of state, elastic properties, and elastic-wave anisotropy can also serve as reference data for geodynamic modeling of the interiors of giant terrestrial planets, particularly when combined with appropriate finite-temperature thermodynamic properties. Beyond the geophysical implications, they provide benchmarks for assessing and validating computational methods [65]. The Landau framework used here can be extended to other ferroelastic rutile-type oxides to investigate their phase-transition mechanisms and associated elastic softening [32,66]. Rutile-type $GeO_2$ is a proposed ultra-wide-bandgap semiconductor for power electronics [67,68]. Its strong elastic anisotropy is relevant to strain engineering in thin films and stress management in electronic devices [69,70].

## Conclusions:

In this study, we have presented a comprehensive first-principles investigation of the structural phase transitions and elastic properties of $GeO_2$ under high pressure, covering the rutile- ($P4_2/mnm$), $CaCl_2$- ($Pnnm$), α-$PbO_2$- ($Pbcn$), and pyrite-type ($Pa\bar{3}$) phases. The rutile-type to $CaCl_2$-type transition is identified as a second-order ferroelastic transition with a critical pressure of 18.3 GPa as determined by strain-based Landau analysis. Using Landau theory to describe the order-parameter–strain coupling, together with the DFT-calculated elastic response, we elucidate the elastic softening near the critical pressure, manifested by the pronounced softening of ($C_{11}$-$C_{12}$) and corresponding anomalies in the bulk and shear moduli. Elastic anisotropy analysis shows that the rutile-type phase exhibits the strongest anisotropy, $A_S$, increasing dramatically from ~71% to ~190% as the transition is approached. Anisotropy decreases sharply in the $CaCl_2$-type phase and becomes nearly constant at ~28% and ~8–12% in the α-$PbO_2$-type and pyrite-type phases, respectively, indicating progressive elastic isotropy under extreme compression. This work provides comprehensive elastic-tensor datasets for four high-pressure polymorphs of $GeO_2$, establishing a quantitative framework for understanding strain-coupling-induced elastic anomalies in rutile-type oxides.

**Acknowledgements:**

We acknowledge the National Supercomputing Mission (NSM) for providing computing resources of 'PARAM Ananta' at IIT Gandhinagar, which is implemented by C-DAC and supported by the Ministry of Electronics and Information Technology (MeitY) and the Department of Science and Technology (DST), Government of India. S.G. thanks CSIR-India for providing the research fellowship. R.D. would like to thank Princeton University for financial support.

**Figures:**

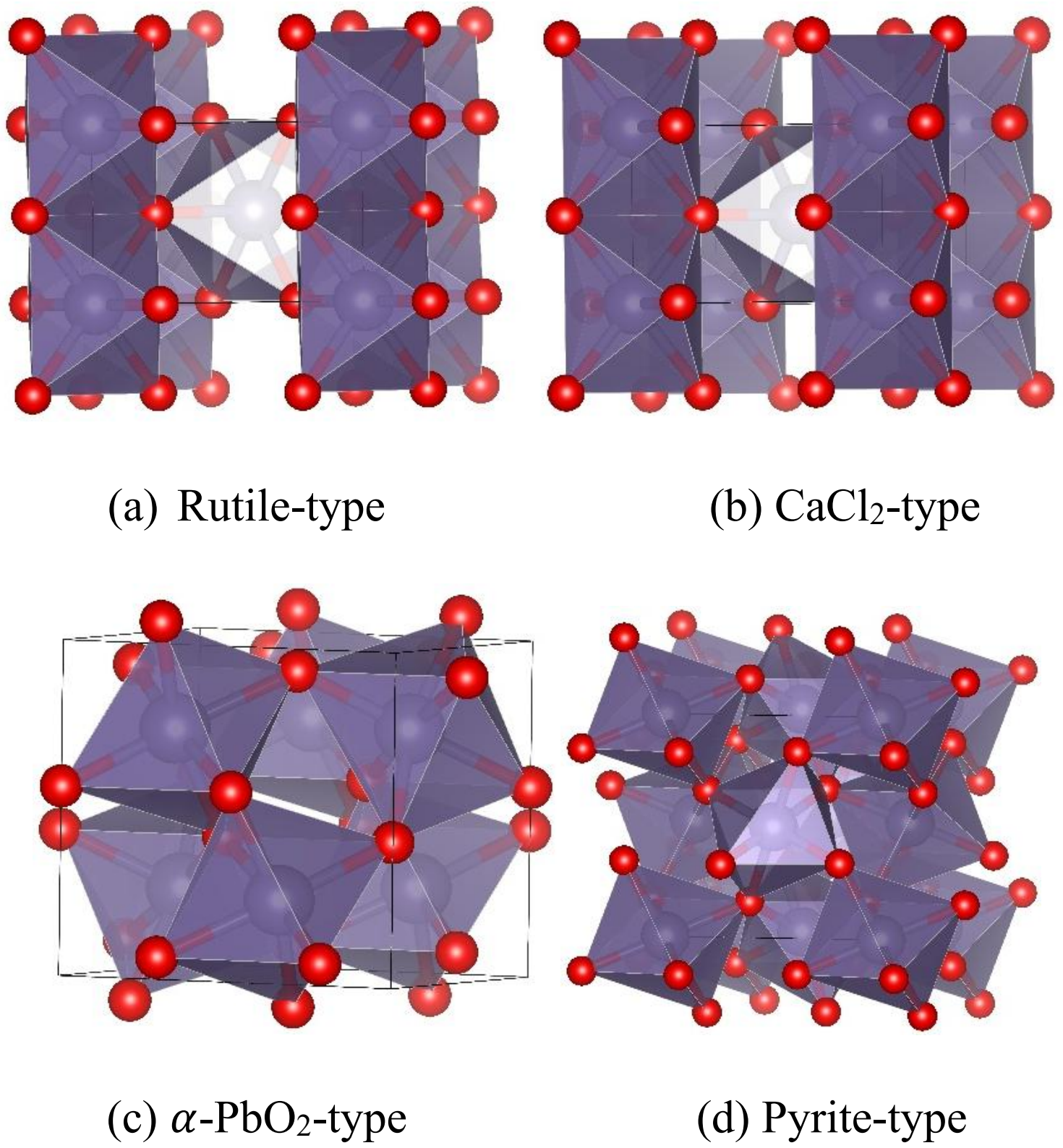

Figure 1. Crystal structures of the four polymorphs of $GeO_2$: (a) rutile-type at 0 GPa, (b) $CaCl_2$-type at 25 GPa, (c) $\alpha$-$PbO_2$-type at 35 GPa, and (d) pyrite-type at 65 GPa. The red and grey spheres represent the oxygen and germanium atoms, respectively.

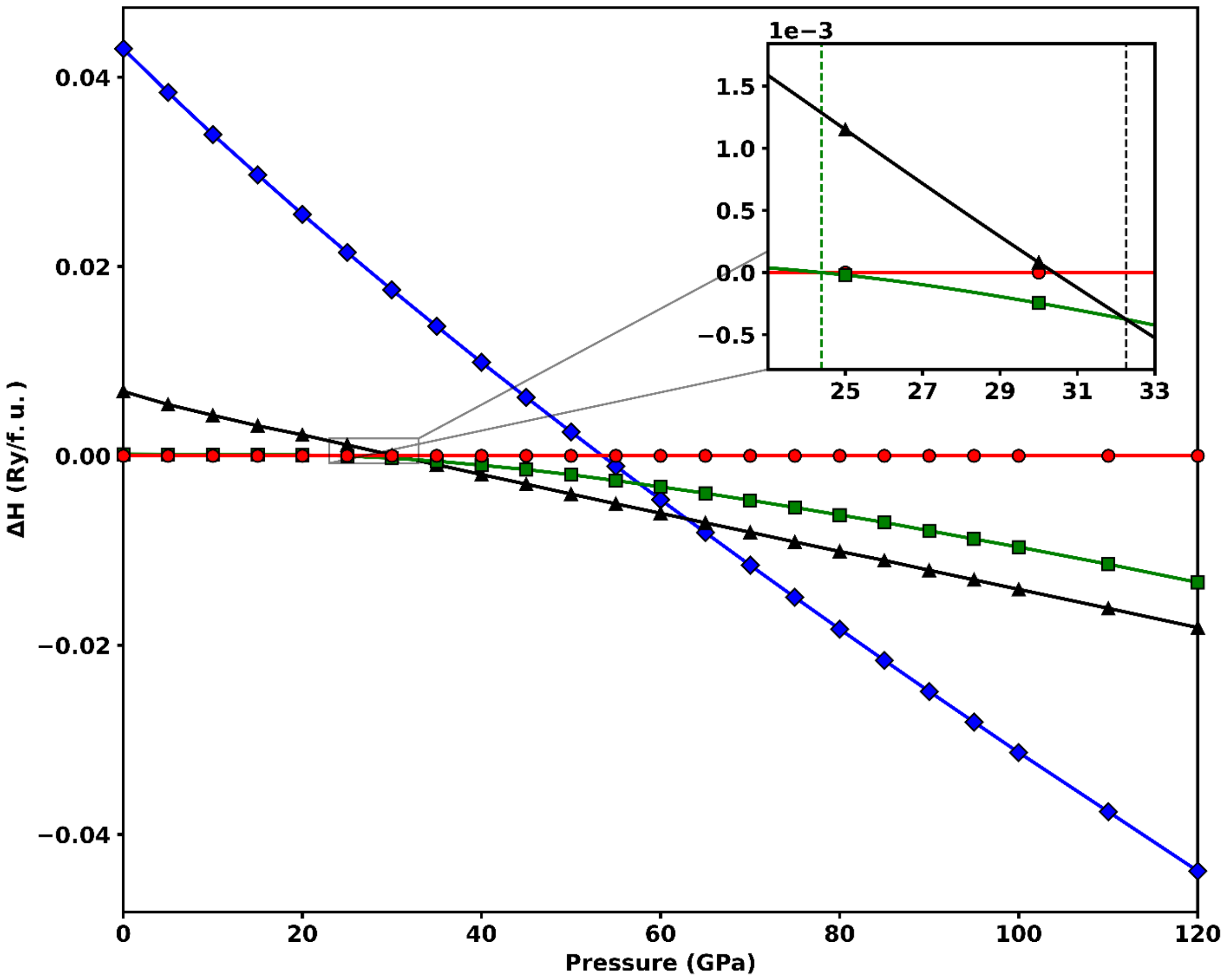

Figure 2. Enthalpy differences of the four $GeO_2$ polymorphs (rutile-type: red circles, $CaCl_2$-type: green squares, α-$PbO_2$-type: black triangles, and pyrite-type: blue vertical squares) relative to the rutile-type phase as a function of pressure. The inset shows an enlarged view of the enthalpy difference in the pressure range of 22.5 to 33 GPa for better visualization.

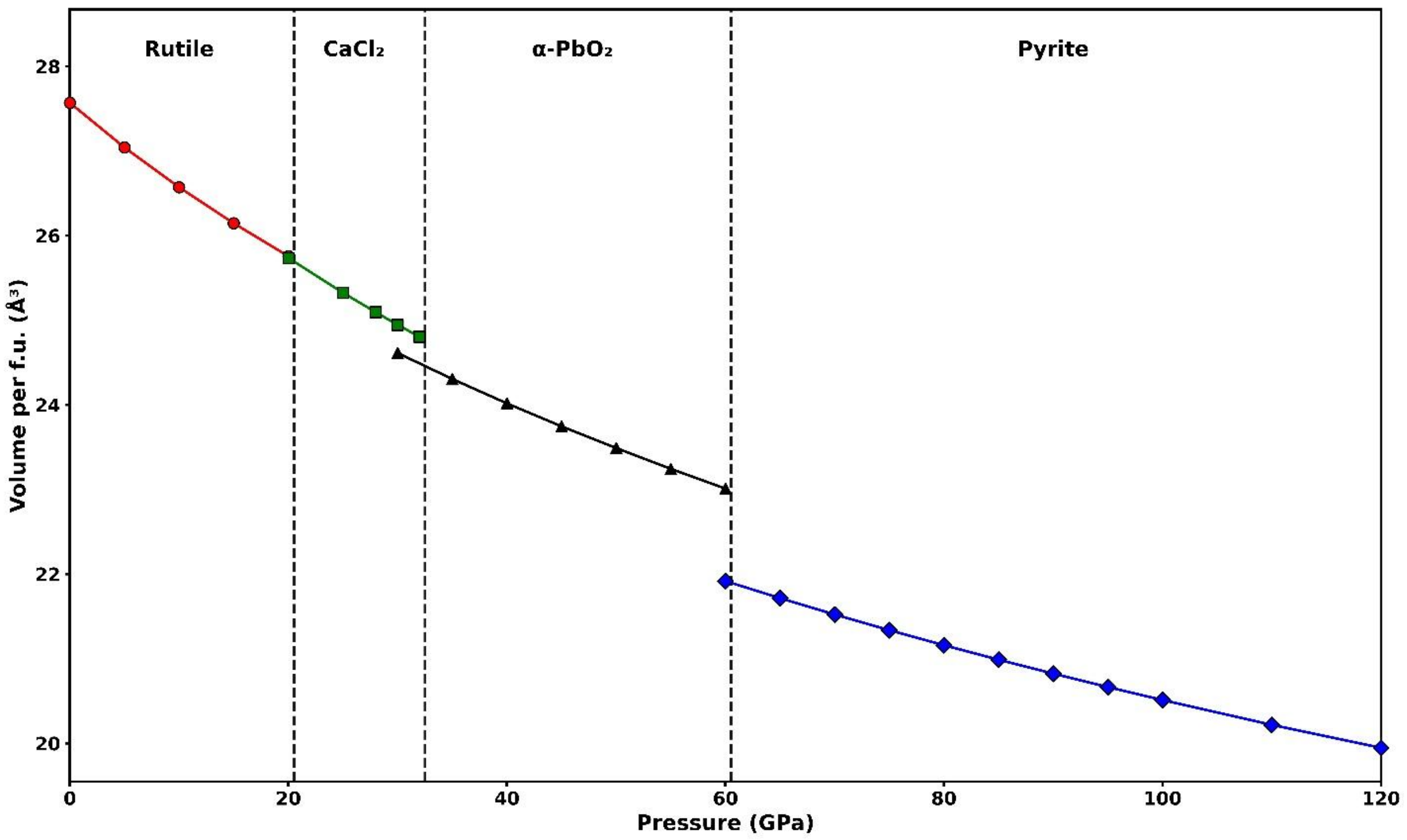

Figure 3. Pressure dependence of the unit cell volume (per formula unit) of the four investigated $GeO_2$ phases. Colors have the same meaning as in Figure 2.

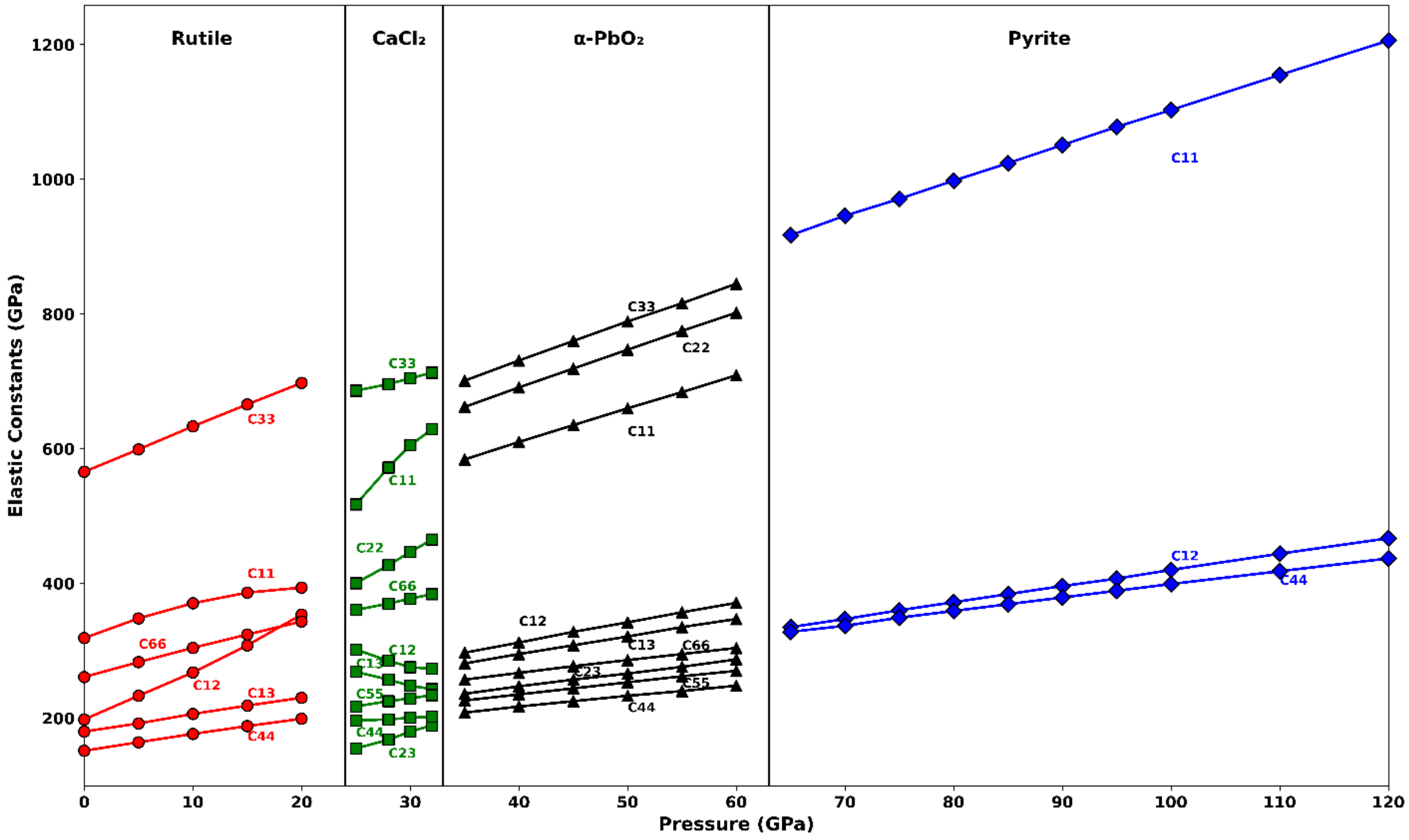

Figure 4. Variation of the elastic constants with pressure for the four $GeO_2$ polymorphs calculated using density functional theory: rutile-type (red), $CaCl_2$-type (green), α-$PbO_2$-type (black), and pyrite-type (blue).

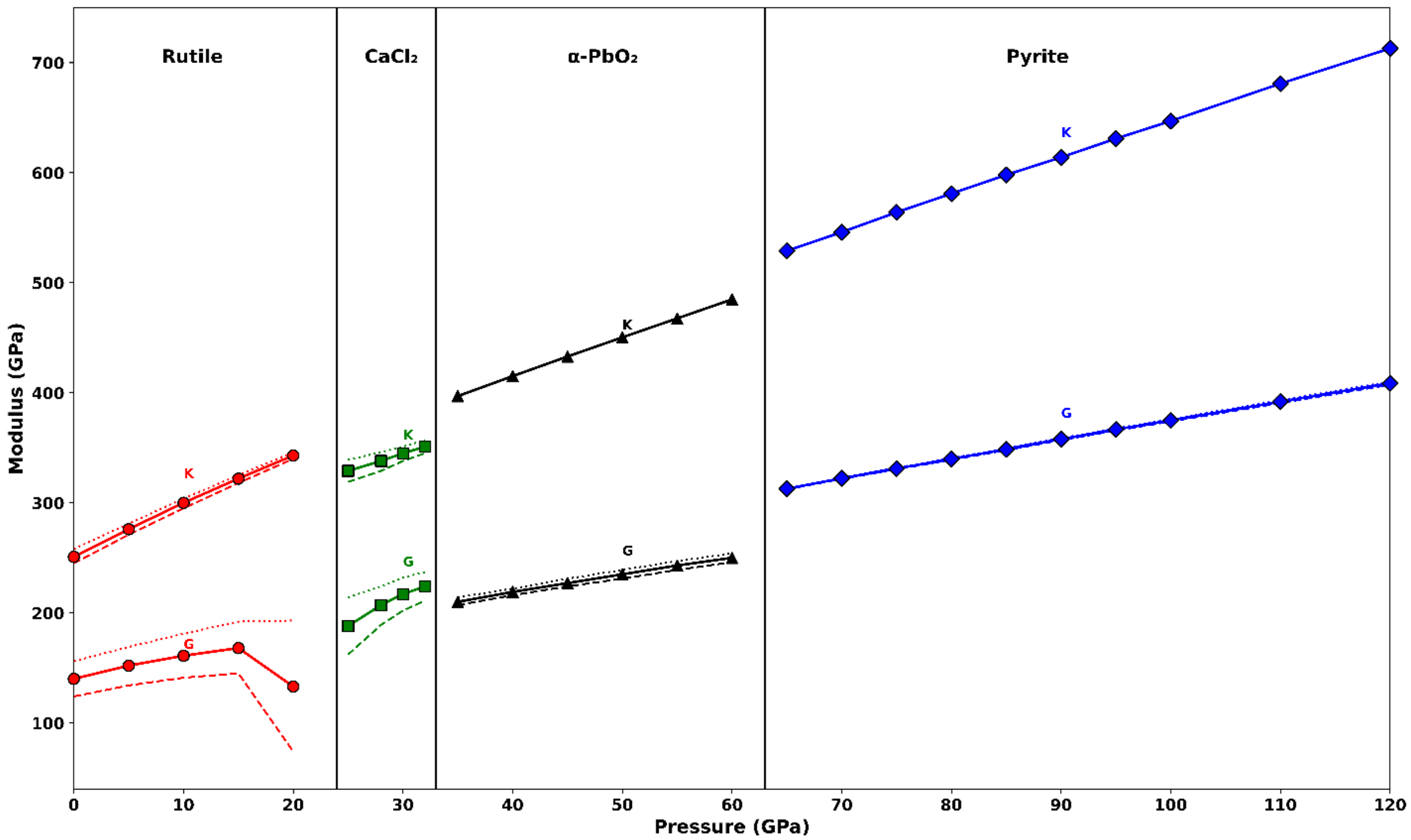

Figure 5. Pressure dependence of the bulk modulus (*K*) and shear modulus (*G*) for the four $GeO_2$ polymorphs calculated using density functional theory. Colors have the same meaning as in Figure 4. The solid lines represent the Hill averages, while the dotted and dashed lines represent the Voigt and Reuss bounds, respectively.

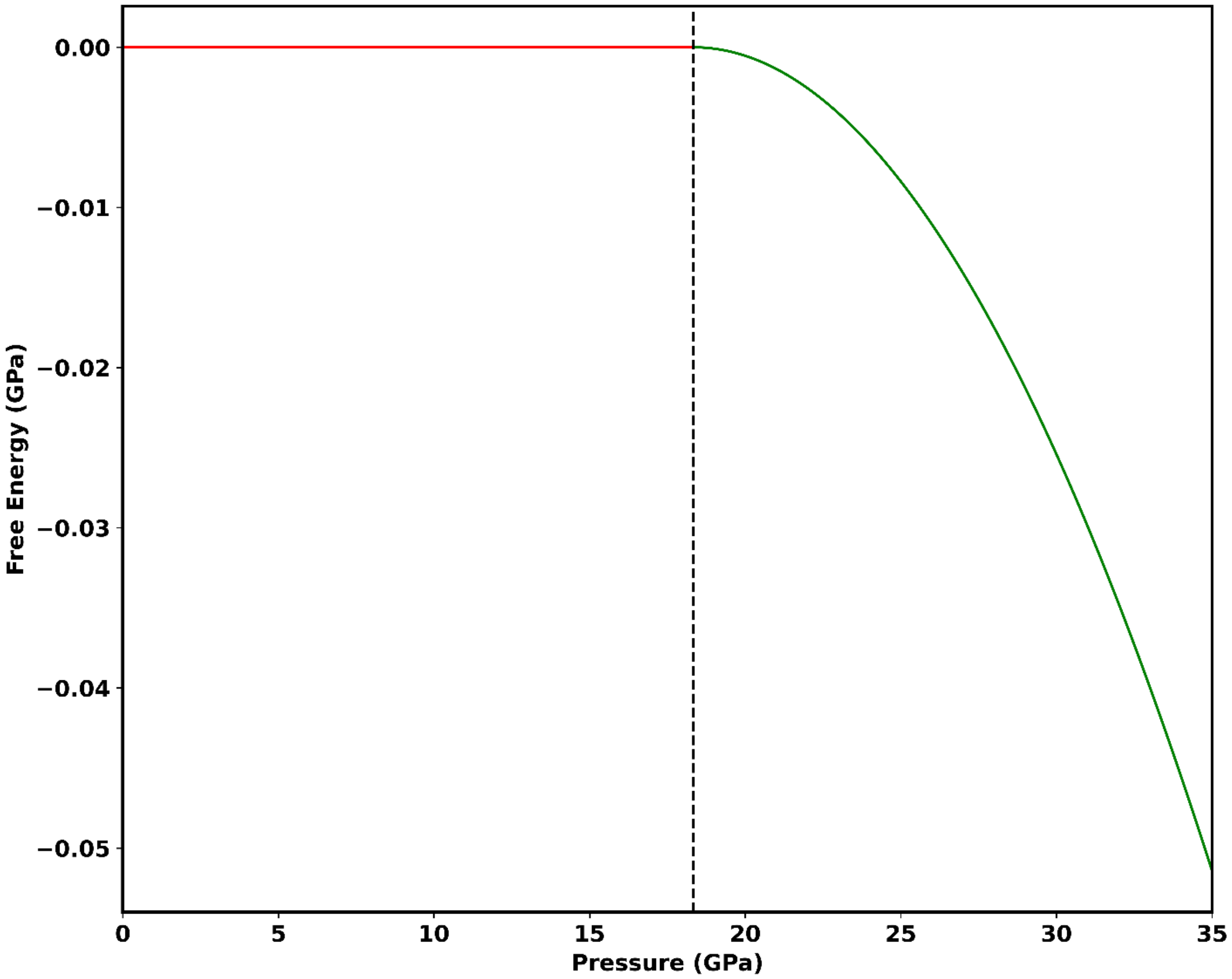

Figure 6. Variation in Landau free energy across the rutile- to $CaCl_2$-type phase transition, illustrating the onset of symmetry-breaking strain and the associated stabilization of the $CaCl_2$-type phase.

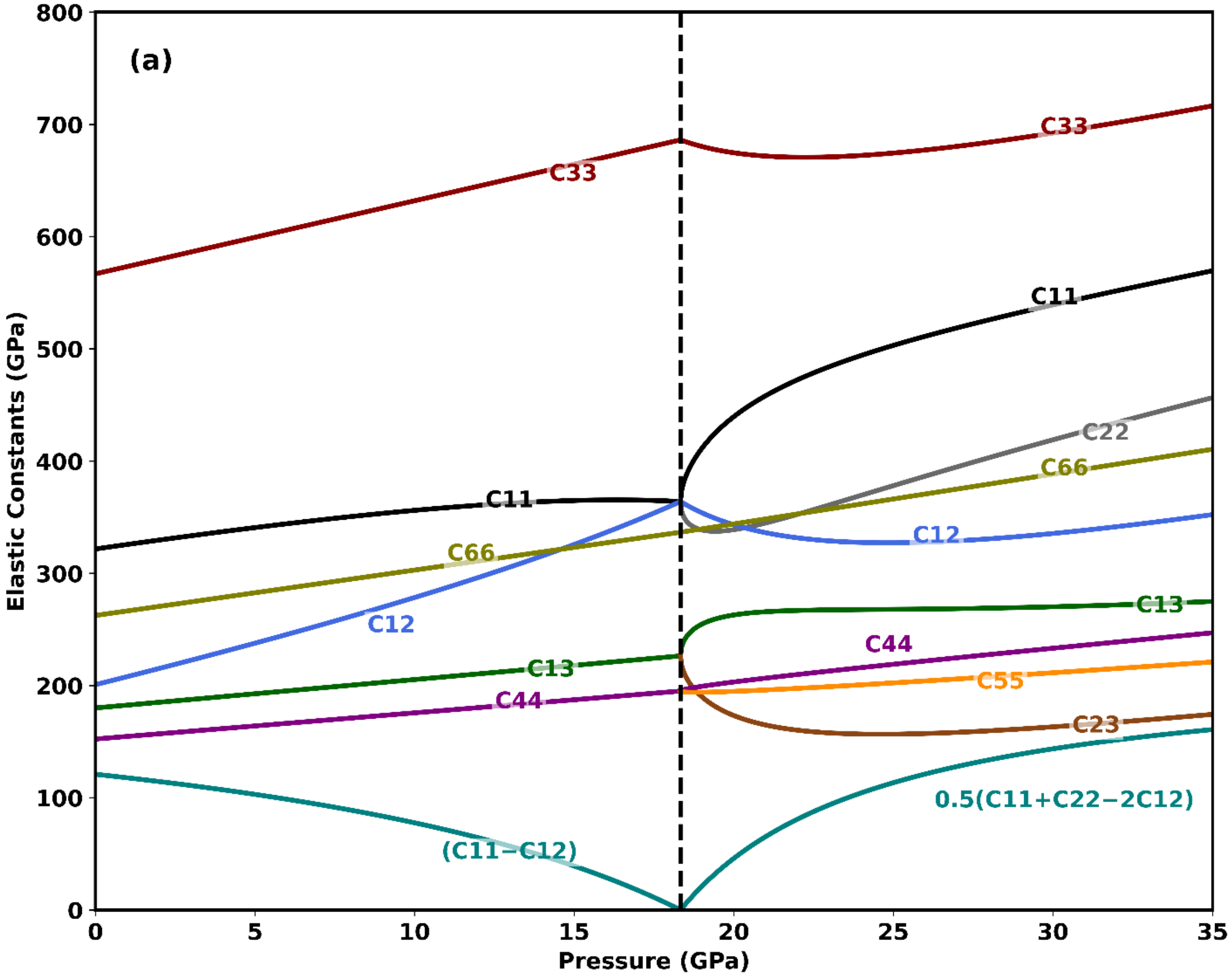

(a)
C33
C33
C11
C11
C22
C66
C12
C66
C12
C13
C13
C44
C44
C55
C23
0.5(C11+C22−2C12)
(C11−C12)
Elastic Constants (GPa)
Pressure (GPa)

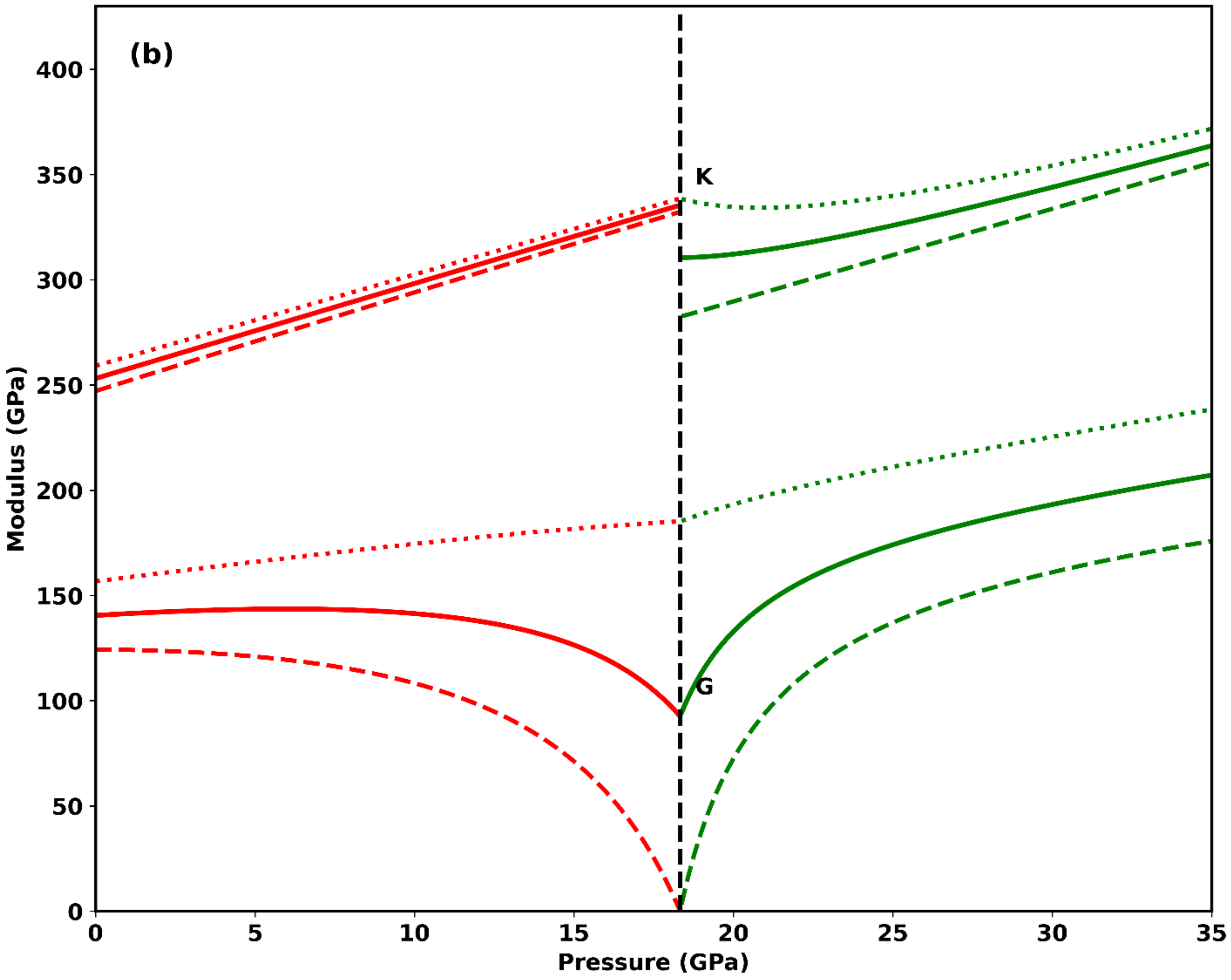

Figure 7. Pressure dependence of the (a) elastic constants and (b) bulk and shear moduli in the rutile- and $CaCl_2$-type phases of $GeO_2$ obtained from the Landau model. In (b), the solid lines represent the Hill averages, while the dotted and dashed lines represent the Voigt and Reuss bounds. Red and green curves correspond to the rutile- and $CaCl_2$-type phases, respectively. The dashed black line shows the critical pressure, $P_S$.

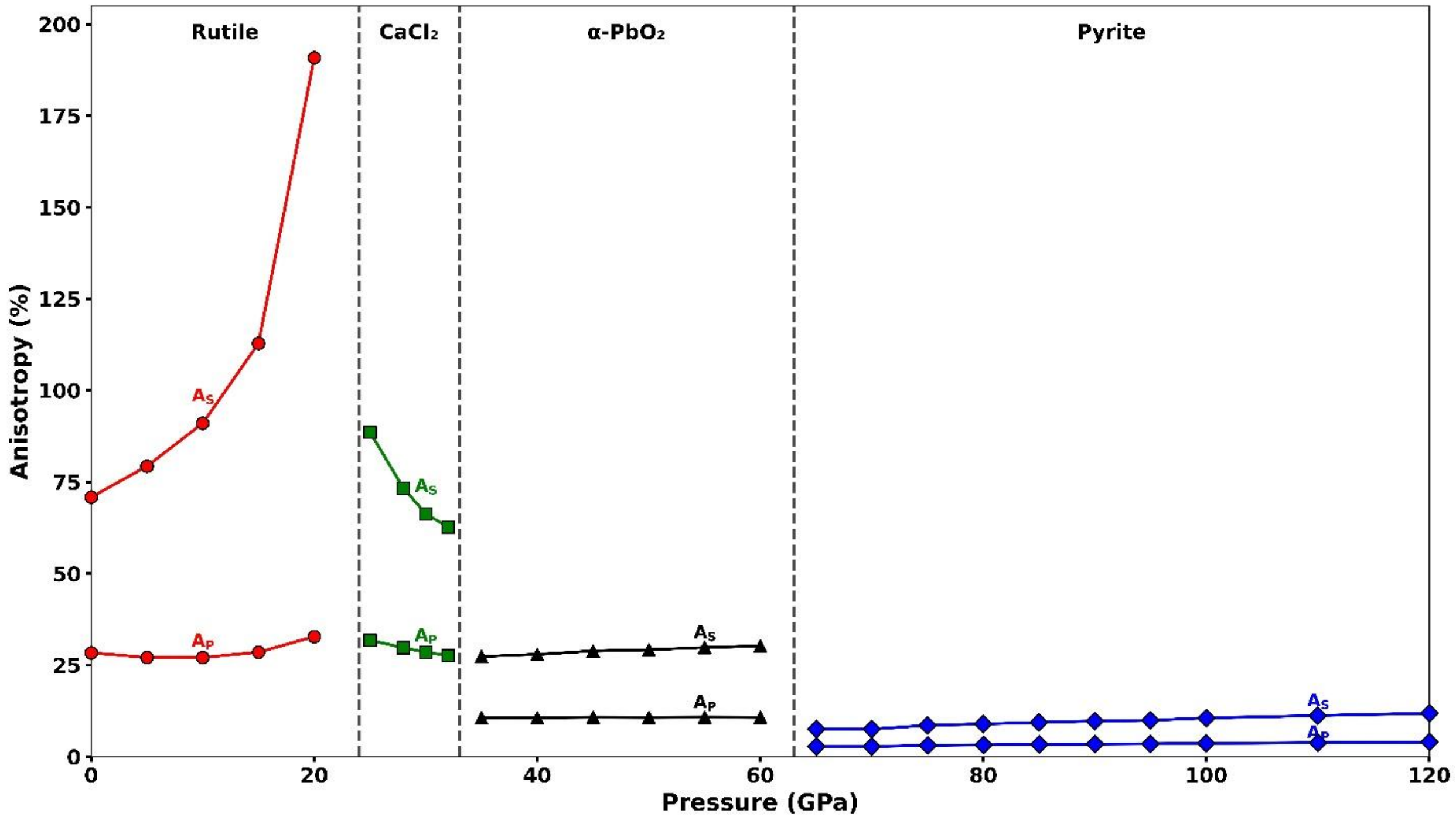

Figure 8. Pressure dependence of the elastic wave anisotropy of the four $GeO_2$ polymorphs calculated using density functional theory; colors have the same meaning as in Figure 4. The anisotropy parameters $A_S$ and $A_P$ represent shear and compressional wave anisotropy, respectively.

**Tables:**

Table I: Comparison of the calculated lattice parameters (Å) and unit-cell volumes ($Å^3$) of $GeO_2$ polymorphs at different pressures with previously reported values ($P_T$ = transition pressure, $Z$ = number of formula units):

| Phase ($GeO_2$) | Method | Pressure (GPa) | $a$(Å) | $b$(Å) | $c$(Å) | $V_0$($Å^3$) | $P_T$ (GPa) |
|---|---|---|---|---|---|---|---|
| Rutile-type | This work | 0 | 4.384 | 2.868 | | 55.12 | |
| (Z = 2) | Exp [71] | 0.0001 | 4.396(4) | 2.863(2) | | 55.33 | |
| | Exp [67] | - | 4.396 | 2.861 | | 55.30 | |
| | Exp [35] | - | 4.406 | 2.861 | | 55.57 | |
| | Exp [72] | - | 4.406(1) | 2.861(2) | | 55.57 | |
| | Hartree-Fock [73] | 0 | 4.352 | 2.863 | | 54.20 | |
| | Empirical pot.[74] | 0 | 4.562 | 2.747 | | 57.20 | |
| $CaCl_2$-type | This work | 25 | 4.298 | 4.184 | 2.816 | 50.65 | 24 |
| (Z = 2) | Exp [46] | 28(1) | 4.284(9) | 4.209(9) | 2.808(3) | 50.63(9) | 28(1) |
| | Exp [28] | 35.9(2) | 4.258(2) | 4.176(1) | 2.796(1) | 49.72(2) | 35.9(2) |
| | Exp [47] | 30.1 | 4.429(7) | 4.403(7) | 2.804(4) | 54.68(7) | 30.1 |
| | Exp [75] | 29.6 | 4.265(1) | 4.225(1) | 2.799(1) | 50.42(1) | 29.6 |
| | DFT-GGA [3] | 30 | 4.375 | 4.134 | 2.844 | 51.44 | 30 |
| α-$PbO_2$-type | This work | 35 | 4.121 | 5.119 | 4.608 | 97.22 | 33 |
| (Z = 4) | Exp [33] | 51.0 | 4.069(1) | 5.074(2) | 4.562(1) | 94.18 | 51.0 |
| | Exp [34] | 60 | 4.042(8) | 5.039(8) | 4.528(0) | 92.25 | 60 |
| | Exp [47] | 70.7 | 4.051(6) | 5.027(6) | 4.522(6) | 92.08 | 70.7 |
| | DFT-LDA [20] | 40 | 4.095 | 5.088 | 4.579 | 95.40 | 40 |
| Pyrite-type | This work | 65 | 4.429 | | | 86.86 | 63 |
| (Z = 4) | Exp [33] | 82.8 | 4.392(1) | | | 84.69 (3) | 82.8 |
| | Exp [47] | 108 | 4.336(15) | | | 81.52 (4) | 108.0 |
| | DFT-LDA [20] | 70 | 4.4046 | | | 85.45 | 70 |

Table II: Comparison of the elastic constants (GPa) and elastic moduli (GPa) of $GeO_2$ polymorphs at 0 K with previously reported theoretical and experimental data:

| Phase | Pressure (GPa) | $C_{11}$ | $C_{22}$ | $C_{33}$ | $C_{44}$ | $C_{55}$ | $C_{66}$ | $C_{12}$ | $C_{13}$ | $C_{23}$ | K | G |
|---|---|---|---|---|---|---|---|---|---|---|---|---|
| Rutile- Type | | | | | | | | | | | | |
| This work | 0 | 318.7 | | 565.7 | 151.4 | | 260.9 | 197.8 | 179.9 | | 251 | 139 |
| Exp [35] | 0.5 | 337.2 | | 599.4 | 161.5 | | 258.4 | 188.2 | 187.4 | | 259 | |
| DFT-LDA [20] | 0 | 316.1 | | 573.9 | 149.9 | | 254.6 | 199.4 | 175.6 | | | |
| DFT-LDA [36] | 0 | 403.5 | | 637.6 | 189.3 | | 285.5 | 184.2 | 188.9 | | 280 | 183 |
| $CaCl_2$-type | | | | | | | | | | | | |
| This work | 30 | 605 | 446.6 | 704.5 | 200.3 | 229.2 | 377.2 | 275.5 | 247.9 | 179.9 | 344 | 217 |
| DFT-LDA [36] | 30 | 587.0 | 414.2 | 728.9 | 208.5 | 230.1 | 388.3 | 369.4 | 232.3 | 156.8 | 356 | 185 |
| $\alpha$-$PbO_2$-type | | | | | | | | | | | | |
| This work | 40 | 610 | 691 | 731 | 217 | 235 | 267 | 312 | 295 | 247 | 415 | 219 |
| DFT-LDA [36] | 40 | 642.3 | 593.8 | 779.7 | 213 | 230 | 287.4 | 359.5 | 334.7 | 305.2 | 444 | 207 |
| Pyrite-type | | | | | | | | | | | | |
| This work | 70 | 946 | | | 337 | | | 347 | | | 546 | 322 |
| DFT-LDA [36] | 70 | 956.6 | | | 304.8 | | | 425.5 | | | 603 | 289 |

Table III. The coefficients of the classical Landau free-energy expansion. $P_S$ is the strain-derived critical pressure, $P_R$ = 19.24 GPa is the pressure at which the $B_{1g}$ and $A_g$ Raman modes cross, and $P_{RC}$ = 35.1 GPa is the pressure at which the extrapolated $B_{1g}$ frequency of the rutile-type phase reaches zero (see supplementary material).

| Bare Elastic Constants | Landau Coefficients |
|---|---|
| $C_{11}^0 = 373.9 + 5.604 \times P$ | $P_S = 18.33$ GPa |
| $C_{33}^0 = 566.8 + 6.512 \times \mathrm{P}$ | $P_{RC} - P_R = 15.86$ GPa |
| $C_{12}^0 = 148.4 + 5.604 \times \mathrm{P}$ | $\lambda_1 = 8.20$ GPa |
| $C_{13}^0 = 180.0 + 2.53 \times \mathrm{P}$ | $\lambda_2 = -9.40$ GPa |
| $C_{11}^0 - C_{12}^0 = 225.5$ | $\lambda_3 = 14.60$ GPa |
| $C_{44}^0 = 152.3 + 2.33 \times \mathrm{P}$ | $a = -0.0494$ |
| $C_{66}^0 = 262.40 + 4.05 \times \mathrm{P}$ | $b = 3.89$ GPa |

## References

[1] Micoulaut M, Cormier L and Henderson G 2006 The structure of amorphous, crystalline and liquid $GeO_2$ J. Phys. Condens. Matter 18 R753-R785. https://doi.org/10.1088/0953-8984/18/45/R01

[2] Ravindra N M, Weeks R A and Kinser D L 1987 Optical properties of $GeO_2$ Phys. Rev. B 36 6132–4. https://doi.org/10.1103/PhysRevB.36.6132

[3] Alptekin S 2020 Phase transition of $GeO_2$ crystal at high pressure: An ab initio molecular dynamics study Int. J. Chem. Technol. 4 90–6. https://doi.org/10.32571/ijct.725601

[4] Shaposhnikov A V, Perevalov T V, Gritsenko V A, Cheng C H and Chin A 2012 Mechanism of $GeO_2$ resistive switching based on the multi-phonon assisted tunneling between traps Appl. Phys. Lett. 100 243506. https://doi.org/10.1063/1.4729589

[5] Kojima K, Tsuchiya K and Wada N 2000 Sol-gel synthesis of $Nd3^{+}$-doped $GeO_2$ glasses and their optical properties J. Sol-Gel Sci. Technol. 19 511–4. https://doi.org/10.1023/A:1008776203454

[6] Trachenko K and Dove M T 2004 Local events and stretched-exponential relaxation in glasses Phys. Rev. B 70 132202. https://doi.org/10.1103/PhysRevB.70.132202

[7] Richet P, Hovis G and Poe B 2004 Energetics of pressure-induced densification in $GeO_2$ glass Chem. Geol. 213 41–7. https://doi.org/10.1016/j.chemgeo.2004.08.031

[8] Sugai S and Onodera A 1996 Medium-range order in permanently densified $SiO_2$ and $GeO_2$ glass Phys. Rev. Lett. 77 4210. https://doi.org/10.1103/PhysRevLett.77.4210

[9] Broqvist P, Binder J F and Pasquarello A 2009 Atomistic model structure of the Ge (1 0 0)–$GeO_2$ interface Microelectron. Eng. 86 1589–91. https://doi.org/10.1016/j.mee.2009.03.087

[10] Da Silva S R M, Rolim G K, Soares G V, Baumvol I J R, Krug C, Miotti L, Freire F L, Da Costa M and Radtke C 2012 Oxygen transport and $GeO_2$ stability during thermal oxidation of Ge Appl. Phys. Lett. 100 191907. https://doi.org/10.1063/1.4712619

[11] Ghobadi E and Capobianco J A 2000 Crystal properties of α-quartz type $GeO_2$ Phys. Chem. Chem. Phys. 2 5761–3. https://doi.org/10.1039/B006839J

[12] Haines J, Cambon O, Philippot E, Chapon L and Hull S 2002 A neutron diffraction study of the thermal stability of the α-quartz-type structure in germanium dioxide J. Solid State Chem. 166 434–41. https://doi.org/10.1006/jssc.2002.9625

[13] Geng Y, Zhang X, Wang F and Liu B 2026 Predict the influence of different defects on the electronic structure and optical properties of two-dimensional $Bi_2S_3$ semiconductor

materials based on first-principles calculations Phys. Scr. 101 195916. https://doi.org/10.1088/1402-4896/ae68c1

[14] Li X, Zhang X, Wang F and Liu B 2025 $Ba_2MNO_6$ (M = Y, Sc; N = Nb, V, Ta) as next-generation double perovskite oxides: a comprehensive first principles study on the stability, mechanical properties, electronic properties and thermophysical properties Phys. Scr. 100 085926. https://doi.org/10.1088/1402-4896/adf0a6

[15] Wu X, Vanderbilt D and Hamann D R 2005 Systematic treatment of displacements, strains, and electric fields in density-functional perturbation theory Phys. Rev. B 72 035105. https://doi.org/10.1103/PhysRevB.72.035105

[16] Nielsen O H and Martin R M 1983 First-principles calculation of stress Phys. Rev. Lett. 50 697–700. https://doi.org/10.1103/PhysRevLett.50.697

[17] Divya R V, Kumar G, Cohen R E, Tracy S J, Meng Y, Chariton S, Prakapenka V B and Dutta R 2024 High-pressure phase transition of olivine-type $Mg_2GeO_4$ to a metastable forsterite-III type structure and their equations of state Am. Mineral. 109 2052–9. https://doi.org/10.2138/am-2023-9208

[18] Baur W H and Khan A A 1971 Rutile-type compounds. IV. $SiO_2$, $GeO_2$ and a comparison with other rutile-type structures Acta Crystallogr. Sect. B 27 2133–9. https://doi.org/10.1107/S0567740871005466

[19] Ono S, Hirose K, Nishiyama N and Isshiki M 2002 Phase boundary between rutile-type and $CaCl_2$-type germanium dioxide determined by in situ X-ray observations Am. Mineral. 87 99–102. https://doi.org/10.2138/am-2002-0110

[20] Łodziana Z, Parlinski K and Hafner J 2001 Ab initio studies of high-pressure transformations in $GeO_2$ Phys. Rev. B 63 134106. https://doi.org/10.1103/PhysRevB.63.134106

[21] Ono S, Tsuchiya T, Hirose K and Ohishi Y 2003 Phase transition between the $CaCl_2$-type and $\alpha$-$PbO_2$-type structures of germanium dioxide Phys. Rev. B 68 134108. https://doi.org/10.1103/PhysRevB.68.134108

[22] Prakapenka V P, Shen G, Dubrovinsky L S, Rivers M L and Sutton S R 2004 High pressure induced phase transformation of $SiO_2$ and $GeO_2$: Difference and similarity J. Phys. Chem. Solids 65 1537–45. https://doi.org/10.1016/j.jpcs.2003.12.019

[23] Dutta R, Kiefer B, Greenberg E, Prakapenka V B and Duffy T S 2019 Ultrahigh-pressure behavior of $AO_2$ (A = Sn, Pb, Hf) compounds J. Phys. Chem. C 123 27735–41. https://doi.org/10.1021/acs.jpcc.9b06856

[24] Shieh S R, Kubo A, Duffy T S, Prakapenka V B and Shen G 2006 High-pressure phases in $SnO_2$ to 117 GPa Phys. Rev. B 73 014105. https://doi.org/10.1103/PhysRevB.73.014105

[25] Kalkan B, Godwal B K, Yan J and Jeanloz R 2022 High-pressure phase transitions and melt structure of $PbO_2$: An analog for silica Phys. Rev. B 105 064111. https://doi.org/10.1103/PhysRevB.105.064111

[26] Grocholski B, Shim S-H, Cottrell E and Prakapenka V B 2014 Crystal structure and compressibility of lead dioxide up to 140 GPa Am. Mineral. 99 170–7. https://doi.org/10.2138/am.2014.4596

[27] Haines J, Léger J, Chateau C, Bini R and Ulivi L 1998 Ferroelastic phase transition in rutile-type germanium dioxide at high pressure Phys. Rev. B 58 R2909. https://doi.org/10.1103/PhysRevB.58.R2909

[28] Ghosh S, Kumar G, Babu Pillai S and Dutta R 2025 Equation of state of the rutile and $CaCl_2$-type phases of $GeO_2$ to 70 GPa J. Appl. Phys. 138 235902. https://doi.org/10.1063/5.0302993

[29] Cohen R E 1992 First-principles predictions of elasticity and phase transitions in high pressure $SiO_2$ and geophysical implications High-Pressure Research: Application to Earth and Planetary Sciences (Geophysical Monograph Series vol 67) ed Y Syono and M H Manghnani (American Geophysical Union) pp 425–31. https://doi.org/10.1029/GM067p0425

[30] Yamada Y, Tsuneyuki S and Matsui Y 1992 Pressure-induced phase transitions in rutile-type crystals High-Pressure Research: Application to Earth and Planetary Sciences (Geophysical Monograph Series vol 67) ed Y Syono and M H Manghnani (American Geophysical Union) pp 441–6. https://doi.org/10.1029/GM067p0441

[31] Karki B B, Stixrude L and Crain J 1997 Ab initio elasticity of three high-pressure polymorphs of silica Geophys. Res. Lett. 24 3269–72. https://doi.org/10.1029/97GL53196

[32] Carpenter M A and Salje E K H 1998 Elastic anomalies in minerals due to structural phase transitions Eur. J. Mineral. 10 693–812. https://doi.org/10.1127/ejm/10/4/0693

[33] Dutta R, White C E, Greenberg E, Prakapenka V B and Duffy T S 2018 Equation of state of the α-$PbO_2$ and $Pa\bar{3}$-type phases of $GeO_2$ to 120 GPa Phys. Rev. B 98 144106. https://doi.org/10.1103/PhysRevB.98.144106

[34] Prakapenka V B, Dubrovinsky L S, Shen G, Rivers M L, Sutton S R, Dmitriev V, Weber H-P and Le Bihan T 2003 α-$PbO_2$-type high-pressure polymorph of $GeO_2$ Phys. Rev. B 67 132101. https://doi.org/10.1103/PhysRevB.67.132101

[35] Wang H and Simmons G 1973 Elasticity of some mantle crystal structures: 2. Rutile $GeO_2$ J. Geophys. Res. 1896-1977 78 1262–73. https://doi.org/10.1029/JB078i008p01262

[36] Liu Q-J and Liu Z-T 2014 Structural, elastic, and mechanical properties of germanium dioxide from first-principles calculations Mater. Sci. Semicond. Process. 27 765–76. https://doi.org/10.1016/j.mssp.2014.07.047

[37] Hohenberg P and Kohn W 1964 Inhomogeneous electron gas Phys. Rev. 136 B864–71. https://doi.org/10.1103/PhysRev.136.B864

[38] Kohn W and Sham L J 1965 Self-consistent equations including exchange and correlation effects Phys. Rev. 140 A1133. https://doi.org/10.1103/PhysRev.140.A1133

[39] Giannozzi P, Baroni S, Bonini N, Calandra M, Car R, Cavazzoni C, Ceresoli D, Chiarotti G L, Cococcioni M, Dabo I, and others 2009 QUANTUM ESPRESSO: a modular and open-source software project for quantum simulations of materials J. Phys. Condens. Matter 21 395502. https://doi.org/10.1088/0953-8984/21/39/395502

[40] Giannozzi P, Andreussi O, Brumme T, Bunau O, Nardelli M B, Calandra M, Car R, Cavazzoni C, Ceresoli D, Cococcioni M, and others 2017 Advanced capabilities for materials modelling with Quantum ESPRESSO J. Phys. Condens. Matter 29 465901. https://doi.org/10.1088/1361-648X/aa8f79

[41] Vanderbilt D 1990 Soft self-consistent pseudopotentials in a generalized eigenvalue formalism Phys. Rev. B 41 7892. https://doi.org/10.1103/PhysRevB.41.7892

[42] Monkhorst H J and Pack J D 1976 Special points for Brillouin-zone integrations Phys. Rev. B 13 5188.5188.https://doi.org/10.1103/PhysRevB.13.5188

[43] Broyden C G 1970 The convergence of a class of double-rank minimization algorithms 1. General considerations IMA J. Appl. Math. 6 76–90. https://doi.org/10.1093/imamat/6.1.76

[44] Golesorkhtabar R, Pavone P, Spitaler J, Puschnig P, and Draxl C 2013. ElaStic: A tool for calculating second-order elastic constants from first principles. Comput. Phys. Commun. 184 1861–73. https://doi.org/10.1016/j.cpc.2013.03.010

[45] Sin'ko G V and Smirnov N A 2002 Ab initio calculations of elastic constants and thermodynamic properties of bcc, fcc, and hcp Al crystals under pressure J. Phys. Condens. Matter 14 6989–7005. https://doi.org/10.1088/0953-8984/14/29/301

[46] Haines J, Leger J M, Chateau C and Pereira A S 2000 Structural evolution of rutile-type and $CaCl_2$-type germanium dioxide at high pressure Phys. Chem. Miner. 27 575–82. https://doi.org/10.1007/s002690000092

[47] Shiraki K, Tsuchiya T and Ono S 2003 Structural refinements of high-pressure phases in germanium dioxide Acta Crystallogr. Sect. B 59 701–8. https://doi.org/10.1107/S0108768103021761

[48] Ono S, Tsuchiya T, Hirose K and Ohishi Y 2003 High-pressure form of pyrite-type germanium dioxide Phys. Rev. B 68 014103. https://doi.org/10.1103/PhysRevB.68.014103

[49] Haines J, Léger J M and Chateau C 2000 Transition to a crystalline high-pressure phase in $\alpha$-$GeO_2$ at room temperature Phys. Rev. B 61 8701–6. https://doi.org/10.1103/PhysRevB.61.8701

[50] Gao J, Liu Q-J and Tang B 2023 Elastic stability criteria of seven crystal systems and their application under pressure: Taking carbon as an example J. Appl. Phys. 133 135901. https://doi.org/10.1063/5.0139232

[51] Carpenter M A, Hemley R J and Mao H 2000 High-pressure elasticity of stishovite and the $P4_2/mnm \rightleftharpoons Pnnm$ phase transition J. Geophys. Res. Solid Earth 105 10807–16. https://doi.org/10.1029/1999JB900419

[52] Bruce A D and Cowley R A 1981 Structural Phase Transitions (London: Taylor & Francis)

[53] Collet E and Azzolina G 2021 Coupling and decoupling of spin crossover and ferroelastic distortion: Unsymmetric hysteresis loop, phase diagram, and sequence of phases Phys. Rev. Mater. 5 044401. https://doi.org/10.1103/PhysRevMaterials.5.044401

[54] Tröster A, Schranz W, Karsai F and Blaha P 2014 Fully consistent finite-strain Landau theory for high-pressure phase transitions Phys. Rev. X 4 031010. https://doi.org/10.1103/PhysRevX.4.031010

[55] Dove M T 1993 Introduction to Lattice Dynamics (Cambridge: Cambridge University Press). https://doi.org/10.1017/CBO9780511619885

[56] Knorr K, Loidl A and Kjems J K 1986 Ferroelastic transition in KBr:KCN studied by neutrons, x-rays and ultrasonic Physica B+C 136 311–14. https://doi.org/10.1016/S0378-4363(86)80080-8

[57] Feile R, Loidl A and Knorr K 1982 Elastic properties of $(KBr)_{(1-x)}$ $(KCN)_x$ Phys. Rev. B 26 6875. https://doi.org/10.1103/PhysRevB.26.6875

[58] Chaikin P M and Lubensky T C 1995 Principles of Condensed Matter Physics (Cambridge: Cambridge University Press). https://doi.org/10.1017/CBO9780511813467

[59] Jiang F, Gwanmesia G D, Dyuzheva T I and Duffy T S 2009 Elasticity of stishovite and acoustic mode softening under high pressure by Brillouin scattering Phys. Earth Planet. Inter. 172 235–40 https://doi.org/10.1016/j.pepi.2008.09.017

[60] Hill R 1952 The elastic behavior of a crystalline aggregate Proc. Phys. Soc. Sect. A 65 349–54 https://doi.org/10.1088/0370-1298/65/5/307

[61] Every A G 1980 General closed-form expressions for acoustic waves in elastically anisotropic solids Physical Review B 22 1746. https://doi.org/10.1103/PhysRevB.22.1746

[62] Mainprice D 1990 A FORTRAN program to calculate seismic anisotropy from the lattice preferred orientation of minerals Comput. Geosci. 16 385–93. https://doi.org/10.1016/0098-3004(90)90072-2

[63] Karki B, Stixrude L, Clark S, Warren M, Ackland G and Crain J 1997 Structure and elasticity of MgO at high pressure Am. Mineral. 82 51–60. https://doi.org/10.2138/am-1997-1-207

[64] Ranganathan S I and Ostoja-Starzewski M 2008 Universal Elastic Anisotropy Index Phys. Rev. Lett. 101 055504. https://doi.org/10.1103/PhysRevLett.101.055504

[65] Lejaeghere K, Bihlmayer G, Björkman T, Blaha P, Blügel S, Blum V, Caliste D et al 2016 Reproducibility in density functional theory calculations of solids Science 351 aad3000. https://doi.org/10.1126/science.aad3000

[66] Myhill R 2025 A model of elastic softening and displacive phase transitions in anisotropic phases, with application to stishovite and post-stishovite Geophys. J. Int. 241 770–96. https://doi.org/10.1093/gji/ggaf043

[67] Galazka Z, Blukis R, Fiedler A, Bin Anooz S, Zhang J, Albrecht M, Remmele T, Schulz T, Klimm D, Pietsch M, Kwasniewski A, Dittmar A, Ganschow S, Juda U, Stolze K, Suendermann M, Schroeder T and Bickermann M 2025 Bulk single crystals and physical properties of rutile $GeO_2$ for high-power electronics and deep-ultraviolet optoelectronics Phys. Status Solidi B 262 2400326. https://doi.org/10.1002/pssb.202400326

[68] Chae S, Lee J, Mengle K A, Heron J T and Kioupakis E 2019 Rutile $GeO_2$: An ultrawide-band-gap semiconductor with ambipolar doping Appl. Phys. Lett. 114 102104. https://doi.org/10.1063/1.5088370

[69] Chae S, Pressley L A, Paik H, Gim J, Werder D, Goodge B H, Kourkoutis L F, Hovden R, McQueen T M, Kioupakis E and Heron J T 2022 Germanium dioxide: A new rutile substrate for epitaxial film growth J. Vac. Sci. Technol. A 40 050401. https://doi.org/10.1116/6.0002011

[70] Chae S, Paik H, Vu N M, Kioupakis E and Heron J T 2020 Epitaxial stabilization of rutile germanium oxide thin film by molecular beam epitaxy Appl. Phys. Lett. 117 072105. https://doi.org/10.1063/5.0018031

[71] Liu L-G, Bassett W A and Sharry J 1978 New high-pressure modifications of $GeO_2$ and $SiO_2$ J. Geophys. Res. Solid Earth 83 2301–5. https://doi.org/10.1029/JB083iB05p02301

[72] Bolzan A A, Fong C, Kennedy B J and Howard C J 1997 Structural studies of rutile-type metal dioxides Acta Crystallogr. Sect. B 53 373–80. https://doi.org/10.1107/S0108768197001468

[73] Jolly L-H, Silvi B and D'arco P 1994 Periodic Hartree-Fock study of minerals: hexacoordinated $SiO_2$ and $GeO_2$ polymorphs Eur. J. Mineral. 6 7–16. https://doi.org/10.1127/ejm/6/1/0007

[74] Oeffner R D and Elliott S R 1998 Interatomic potential for germanium dioxide empirically fitted to an ab initio energy surface Phys. Rev. B 58 14791–803. https://doi.org/10.1103/PhysRevB.58.14791

[75] Smith G A, Schacher D, Hinton J K, Sneed D, Park C, Petitgirard S, Lawler K V and Salamat A 2021 Prevalence of pretransition disordering in the rutile-to-$CaCl_2$-type phase transition of $GeO_2$ Phys. Rev. B 104 134107. https://doi.org/10.1103/PhysRevB.104.134107

**Supplementary Material**

*S1. Landau theory:*

The Landau Gibbs free energy expansion for a ferroelastic transition with a single order parameter coupled with strain is as follows [1].

$$G = \frac{1}{2}a(P - P_{RC})Q^2 + \frac{1}{4}bQ^4 + \lambda_1(e_1 + e_2)Q^2 + \lambda_2(e_1 - e_2)Q$$
$$+\lambda_3 e_3 Q^2 + \lambda_4(e_4^2 - e_5^2)Q + \lambda_6 e_6^2 Q^2$$
$$+\frac{1}{4}(C_{11}^0 + C_{12}^0)(e_1 + e_2)^2 + \frac{1}{4}(C_{11}^0 - C_{12}^0)(e_1 - e_2)^2$$
$$+C_{13}^0(e_1 + e_2)e_3 + \frac{1}{2}C_{33}^0 e_3^2 + \frac{1}{2}C_{44}^0(e_4^2 + e_5^2) + \frac{1}{2}C_{66}^0 e_6^2 \quad \text{(S1)}$$

Here, $G$ represents the Landau free energy, with $Q$ as an order parameter, while $a$ and $b$ are Landau-normalized coefficients, $e_i$ $(i = 1-6)$ are the spontaneous strains, $\lambda_i$ $(i = 1-6)$ are the order coupling parameter coefficients and $C_{ij}^0$ are the bare elastic constants, which are unaffected by the phase transition.

The displacive phase transition from strain arrived at 18.33 GPa ($P_s$), while $P_{RC}$ is the pressure at which Raman modes go to zero for the rutile phase (see Section S3). The normalized equations are written as:

$$P_R = P_{RC} + \frac{\lambda_2^2}{0.5a(C_{11}^0 - C_{12}^0)} \quad \text{(S2)}$$

The order parameter is related to pressure as:

$$Q^2 = \frac{a(P_S - P)}{b^*} \quad \text{(S3)}$$

The free energy can be rewritten in reduced form as:

$$G = 0.5\,a(P - P_s)Q^2 + 0.25b^*Q^4 \quad \text{(S4)}$$

The effective quadratic coefficient is given by:

$$b^* = b - 2\left[\frac{\lambda_3^2(C_{11}^0 + C_{12}^0) + 2\,\lambda_1^2 C_{33}^0 - 4\lambda_1\lambda_3 C_{13}^0}{(C_{11}^0 + C_{12}^0)C_{33}^0 - 2{C_{13}^0}^2}\right] \quad \text{(S5)}$$

At equilibrium, the strain components vary with Q as:

$$e_1 - e_2 = \frac{-2\lambda_2 Q}{(C_{11}^0 - C_{12}^0)} \quad \text{(S6)}$$

$$e_1 + e_2 = \left(\frac{2\lambda_3 C_{13}^0 - 2\lambda_1 C_{33}^0}{C_{33}^0[C_{11}^0 + C_{12}^0] - 2(C_{13}^0)^2}\right)Q^2 \quad \text{(S7)}$$

*Corresponding author: Gulshan Kumar (kumargulshan@iitgn.ac.in)

$$e_3 = \left(\frac{2\lambda_1 C_{33}^0 - \lambda_3[C_{11}^0 + C_{12}^0]}{C_{33}^0[C_{11}^0 + C_{12}^0] - 2(C_{13}^0)^2}\right) Q^2 \quad \text{(S8)}$$

Strains $e_4$, $e_5$, and $e_6$ remain zero in the orthorhombic phase.

The inverse susceptibility of the order parameter is:

$$\chi^{-1} = a(P - P_{RC}), \quad for\ P < P_s \quad \text{(S9)}$$

$$\chi^{-1} = \frac{(3ab - a)(P_s - P)}{b^*} + a(P_s - P_{RC}), for\ P > P_s \quad \text{(S10)}$$

Elastic anomalies across the transition can be derived from second derivatives of G with respect to strain components [2].

$$C_{ik} = C_{ik}^0 - \Sigma \frac{\partial^2 G}{\partial e_i \partial Q} \left(\frac{\partial^2 G}{\partial Q^2}\right)^{-1} \frac{\partial^2 G}{\partial e_k \partial Q} \quad \text{(S11)}$$

The relevant elastic constant combination is expressed in terms of the pressure deviation from the transition pressure [2–4] as:

$$C_{11} - C_{12} = (C_{11}^0 - C_{12}^0) \frac{(P - P_S)}{(P - (P_S + P_{RC} - P_R))} \quad \text{(S12)}$$

The relationship between the susceptibility and the squared frequency can be illustrated using the potential energy of a harmonic oscillator in a double-well potential [5–7]. Assuming the Landau expansion near the equilibrium position:

$$G(Q) \approx G(Q_0) + 0.5 \frac{\delta^2 G}{\delta Q^2}|_{Q_0}(Q - Q_0)^2 \quad \text{(S13)}$$

And the curvature here for the free energy acts like an effective spring constant (k):

$$\text{Curvature} = k = \frac{\delta^2 G}{\delta Q^2}|_{Q_0} \quad \text{(S14)}$$

For a soft mode, the effective mass of the atom M can be written as:

$$M \frac{\partial^2 Q}{\partial t^2} = -\frac{\partial G}{\partial Q} \quad \text{(S15)}$$

This is the general equation for a harmonic oscillator, which gives:

$$\omega^2 \propto \frac{k}{M} \quad \text{(S16)}$$

This is the inverse of the susceptibility given by S9 and S10.

*S2. Strain Analysis:*

The spontaneous strains $e_1$, $e_2$, $e_3$ for a tetragonal to orthorhombic transition are defined as:

$$e_1 = \frac{(a - a_0)}{a_0} \quad \text{(S17)}$$

$$e_2 = \frac{(b - a_0)}{a_0} \quad \text{(S18)}$$

$$e_3 = \frac{(c - c_0)}{c_0} \quad \text{(S19)}$$

where $a, b,$ and $c$ are lattice parameters of the orthorhombic $CaCl_2$-type phase; $a_0$ and $c_0$ are the lattice parameters of the tetragonal rutile-type phase extrapolated into the stability field of the $CaCl_2$-type phase.

The symmetry-breaking strain is:

$$(e_1 - e_2) = \frac{(a - b)}{a_0} \quad \text{(S20)}$$

If the non-symmetry-breaking strain $(e_1 + e_2)$ is small, we can approximate the reference lattice constant as:

$$a_0 = sqrt(a * b) \quad \text{(S21)}$$

where the volume strain is expressed as:

$$v_s = \frac{(V - V_0)}{V_0} \quad \text{(S22)}$$

where V and $V_0$ are the unit-cell volumes of the orthorhombic $CaCl_2$-type phase and the extrapolated tetragonal rutile-type phase, respectively:

$$v_s \approx e_1 + e_2 + e_3 \quad \text{(S23)}$$

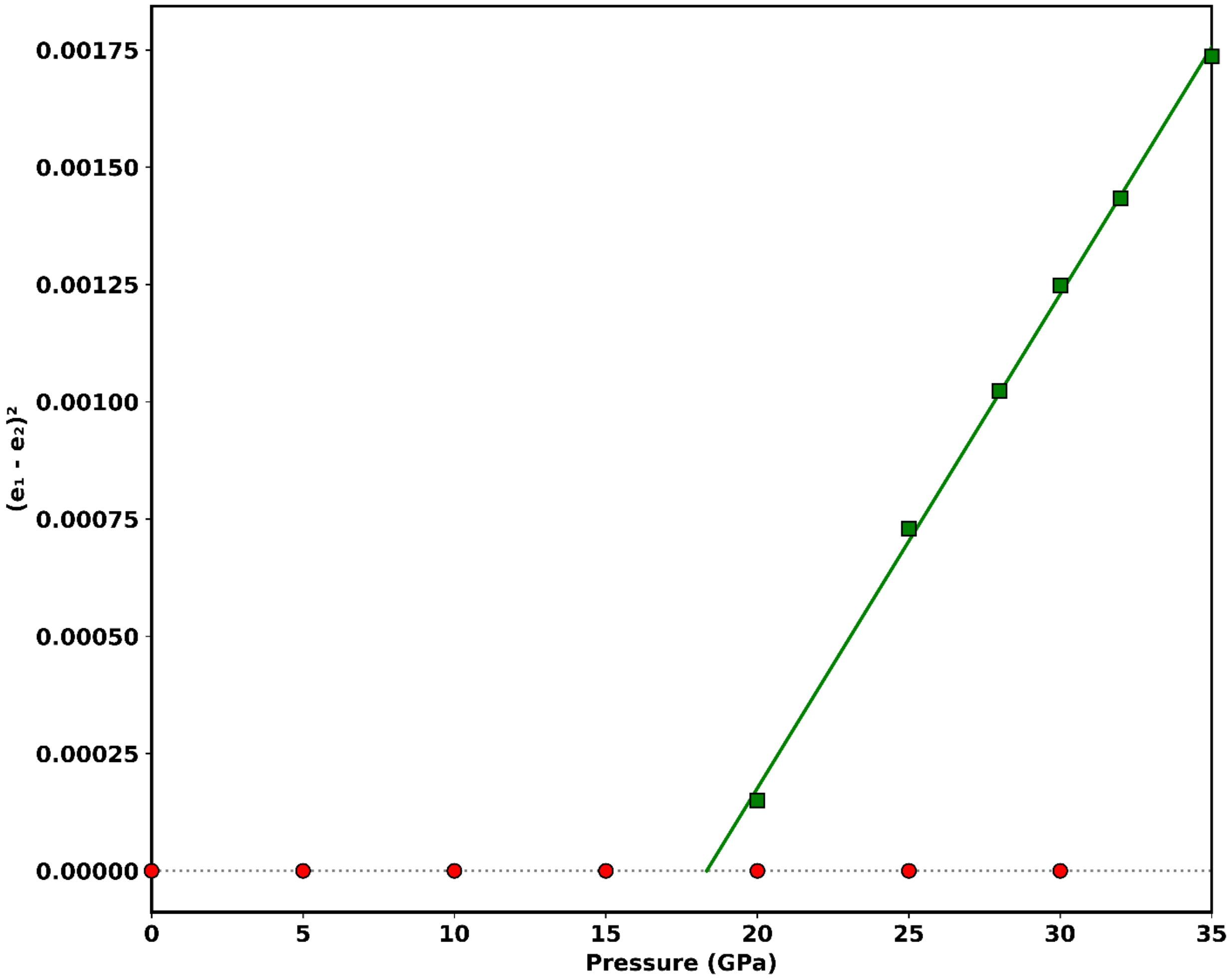

Figure S1: Pressure dependence of the strain difference in the rutile- and $CaCl_2$-type phases. The rutile-type phase (red circles) shows zero strain difference, whereas the $CaCl_2$-type phase (green squares) shows a linear increase in strain difference with increasing pressure.

*S3. Critical pressure calculated from phonon calculations:*

The pressure dependence of the critical Raman modes of the rutile- and $CaCl_2$-type phases, as reported in Ref. [8], can be expressed as:

$$B_{1g}\text{(Rutile-type phase)} = -637 \cdot P + 22358\ cm^{-2}$$

$$A_g\ (\text{CaCl}_2\text{-type Phase}) = 1557 \cdot P - 19855\ cm^{-2}$$

The second-order phase-transition pressure from both Raman modes (Ag and B1g) crossing ($P_R$) and the corresponding soft-mode critical pressure from extrapolating the Raman mode ($B_{1g}$) for rutile phases, where it goes to 0 ($P_{CR}$),[6,7] are 19.24 GPa and 35.1 GPa, respectively, as shown in Figure S2.

The slope ratio of the above equation is $\frac{|slope_Ag|}{|slope_B_1g|} = \frac{1557}{637} \approx$ 2.44 greater than the ideal value of 2 [1]. This deviation suggests that ($e_3$ and ($e_1 + e_2$)) make a finite contribution to the free energy and constrain the strain coupling coefficients $\lambda_1$ and $\lambda_3$.

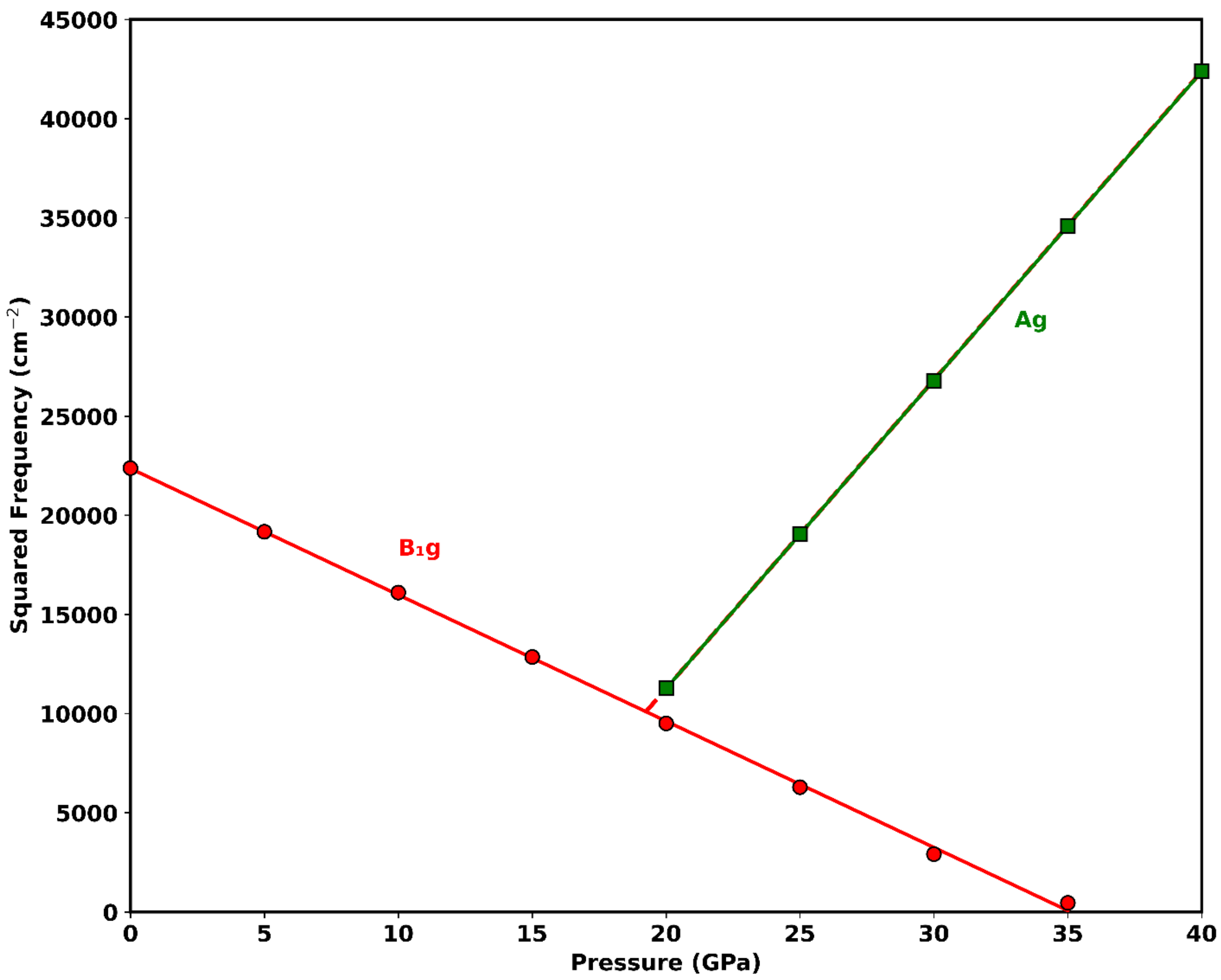

Figure S2: Pressure dependence of the critical Raman modes (modified after Ref. 8); $B_{1g}$ for the rutile-type (red circles) and $A_g$ (green squares) for the $CaCl_2$-type phase.

*S4. Shear and Bulk modulus calculations:*

We calculated the bulk (*K*) and shear (*G*) moduli using the following general expressions [9]:

$$K_R = \frac{1}{(S_{11} + S_{22} + S_{33}) + 2(S_{12} + S_{13} + S_{23})} \quad \text{(S24)}$$

$$K_V = \frac{1}{9}(C_{11} + C_{22} + C_{33}) + \frac{2}{9}(C_{12} + C_{13} + C_{23}) \quad \text{(S25)}$$

$$G_R = \frac{15}{4(S_{11}+S_{22}+S_{33}) - 4(S_{12}+S_{13}+S_{23}) + 3(S_{44}+S_{55}+S_{66})} \quad \text{(S26)}$$

$$G_V = \frac{1}{15}(C_{11} + C_{22} + C_{33} - C_{12} - C_{13} - C_{23}) + \frac{1}{5}(C_{44} + C_{55} + C_{66}) \quad \text{(S27)}$$

The subscripts R and V represent the Reuss and Voigt bounds of the elastic moduli. $S_{ij}$ represents the elastic compliance tensor ($S_{ij} = 1/C_{ij}$). The isotropic bulk (*K*) and shear (*G*) moduli are calculated using Hill's average, defined as the arithmetic mean of the Reuss and Voigt bounds:

$$K = \frac{1}{2}(K_R + K_V) \text{ an } G = \frac{1}{2}(G_R + G_V) \quad \text{(S28)}$$

Using *K*, *G,* and the density ($\rho$) of the crystal, we calculate the compressional ($V_P$) and shear ($V_S$) elastic wave velocities as:

$$V_P = \sqrt{\frac{K + \frac{4}{3}G}{\rho}} \text{ and } V_S = \sqrt{\frac{G}{\rho}} \quad \text{(S29)}$$

**References:**

[1] Carpenter M A, Hemley R J and Mao H 2000 High-pressure elasticity of stishovite and the $P4_2/mnm \rightleftharpoons Pnnm$ phase transition. J. Geophys. Res. Solid Earth **105** 10807–16 https://doi.org/10.1029/1999JB900419

[2] Knorr K, Loidl A and Kjems J K 1986 Ferroelastic transition in KBr: KCN studied by neutrons, X-rays and ultrasonic Physica B+C 136 311–14. https://doi.org/10.1016/S0378-4363(86)80080-8

[3] Carpenter M A and Salje E K H 1998 Elastic anomalies in minerals due to structural phase transitions Eur. J. Mineral. 10 693–812. https://doi.org/10.1127/ejm/10/4/0693

[4] Feile R, Loidl A and Knorr K 1982 Elastic properties of $(KBr)_{1-x}(KCN)_x$ Phys. Rev. B 26 6875. https://doi.org/10.1103/PhysRevB.26.6875

[5] Chaikin P M and Lubensky T C 1995 Principles of Condensed Matter Physics (Cambridge: Cambridge University Press). https://doi.org/10.1017/CBO9780511813467

[6] Bruce A D and Cowley R A 1981 Structural Phase Transitions (London: Taylor & Francis).

[7] Dove M T 1993 Introduction to Lattice Dynamics (Cambridge: Cambridge University Press). https://doi.org/10.1017/CBO9780511619885

[8] Ghosh S, Kumar G, Babu Pillai S and Dutta R 2025 Equation of state of the rutile and $CaCl_2$-type phases of $GeO_2$ to 70 GPa J. Appl. Phys. 138 235902. https://doi.org/10.1063/5.0302993

[9] Hill R 1952 The elastic behaviour of a crystalline aggregate Proc. Phys. Soc. Sect. A 65 349. https://doi.org/10.1088/0370-1298/65/5/307